\documentclass[aps,superscriptaddress,amsmath,amssymb,twocolumn,amsfonts,floatfix, longbibliography]{revtex4-2}
\usepackage{mathtools}
\usepackage{braket}
\usepackage[dvipsnames]{xcolor}
\usepackage{float}
\usepackage{upgreek}
\usepackage{pgffor}
\usepackage{float}
\usepackage{silence}
\usepackage[colorlinks=true,linktoc=page,linkcolor=blue,citecolor=magenta,urlcolor=Fuchsia]{hyperref}
\usepackage[normalem]{ulem}
\usepackage{sidecap,tikz}
\usepackage{orcidlink}

\newcommand{\beq}{\begin{equation}}
\newcommand{\eeq}{\end{equation}}
\newcommand{\bea}{\begin{eqnarray}}
\newcommand{\eea}{\end{eqnarray}}

\newcommand{\eqn}[1] {Eq.~(\ref{#1})}
\newcommand{\fig}[1]{Fig.~\ref{#1}}
\newcommand{\sect}[1]{Sec.~\ref{#1}}
\newcommand{\mylabel}[1]{\label{#1}}

\mathchardef\mhyphen="2D 
\newcommand{\noin}{\noindent}

\newcommand{\ua}{{\uparrow }}
\newcommand{\da}{{\downarrow }}

\newcommand{\non}{\nonumber}

\begin{document}
\title{Superconducting diode effect in multichannel Majorana wires}	

\author{{Sagar Santra}}
\thanks{These authors contributed equally to this work}
\affiliation{Department of Physics, Indian Institute of Technology, Kanpur 208016, India}

\author{{Dibyendu Samanta}\,\orcidlink{0009-0004-3022-7633}}
\thanks{These authors contributed equally to this work}
\affiliation{Department of Physics, Indian Institute of Technology, Kanpur 208016, India}

\author{{Sudeep Kumar Ghosh}\,\orcidlink{0000-0002-3646-0629}}
\email{skghosh@iitk.ac.in}
\affiliation{Department of Physics, Indian Institute of Technology, Kanpur 208016, India}
	
\date{\today}

\begin{abstract}
The superconducting diode effect (SDE) enables nonreciprocal dissipationless transport when inversion and time-reversal symmetries are simultaneously broken. Rashba nanowires proximitized by conventional $s$-wave superconductors provide a minimal setting where spin–orbit coupling and Zeeman fields generate asymmetric finite-momentum pairing. While most studies focus on the single-channel limit, which typically yields small diode efficiencies and requires multiple Zeeman-field components, realistic devices generically host multiple transverse subbands (channels). Here, we investigate the SDE in multichannel Rashba nanowires with harmonic and rectangular quantum-well confinement using a self-consistent Bogoliubov–de Gennes formalism. Both geometries support asymmetric Fulde–Ferrell (FF) states driving pronounced nonreciprocal supercurrents. Crucially, this current-driven FF state stabilizes a topological phase with Majorana zero modes, where Cooper pair momentum is controlled by an externally injected supercurrent, enabling direct topological manipulation. Pairing susceptibility analysis reveals that field-induced asymmetries favor directional Cooper pairing, explaining the diode response's nonmonotonic Zeeman-field dependence. Harmonic confinement yields diode efficiencies of $\sim 60\%$ (interacting channels) and $\sim 55\%$ (independent channels). Notably, interchannel coupling enables a finite response from a transverse Zeeman field alone. Rectangular confinement achieves $\sim 60\%$ efficiency across both regimes, alongside a tunable sign reversal of efficiency when channels interact. These results establish the robustness of the SDE and FF states against transverse confinement variations, highlighting multichannel nanowires as powerful platforms for high-efficiency nonreciprocal transport and current-controlled topological superconductivity.

\end{abstract}
	
\maketitle

\section{Introduction}
The discovery of nonreciprocal supercurrent transport in superconductors has opened a new frontier in superconducting electronics, positioning the superconducting diode effect (SDE) as a central paradigm for next-generation dissipationless devices~\cite{Ando2020,Daido_2022}. In the SDE, the critical supercurrent depends on the direction of current flow, producing diode-like behavior without energy dissipation~\cite{nadeem_2023,Ma2025,shaffer2025}. From a symmetry standpoint, such nonreciprocity is forbidden in conventional superconductors and emerges only when inversion and time-reversal symmetries are simultaneously broken. While this condition can be achieved through finite-momentum pairing states, it may also arise intrinsically in noncentrosymmetric superconductors via spontaneous time-reversal symmetry breaking, providing an alternative route to nonreciprocal supercurrent transport~\cite{Ghosh2020a,Shang2022Weyl,Shang2018,Shang2020,Sajilesh2025time,kataria2026}. Guided by this principle, extensive theoretical efforts have proposed mechanisms for realizing the SDE in both tunnel-junction-based architectures~\cite{Zhang_2022,Kokkeler_2022,Tanaka_2022,Souto_2022,Cheng_2023,Steiner_2023,Costa_2023,Wei_2022} and junction-free superconducting phases~\cite{Daido_2022,akito_2022,Yuan_2022,He_2022,ilic_2022,Scammell_2022,Zinkl_2022,He_2023,Zhai_2022,Jiang_2022,Picoli_2023,Legg_2022,Banerjee_2024,bhowmik_2025,dibyendu_2025,bhowmik2025,Sayak2025,schrade2026,Amartya2025,ruthvik2025}, using approaches ranging from phenomenological Ginzburg–Landau theory to fully microscopic mean-field treatments. Finite-momentum superconductivity provides a particularly transparent route to nonreciprocity, with asymmetric Fulde–Ferrell (FF) pairing offering a minimal realization in which both inversion and time-reversal symmetries are effectively broken~\cite{Yuan_2022,Fulde1964}. Experimentally, nonreciprocal superconductivity has now been demonstrated across diverse material platforms, including artificial superlattices~\cite{Narita_2024,sundaresh_2023}, bulk materials and thin films~\cite{Wakatsuki_2017,nadeem_2023,Yuki_2020,Schumann_2020}, multilayer graphene systems~\cite{lin_2022,Jaime_2023,Chen2025}, and transition-metal dichalcogenides~\cite{bauriedl_2022,Yun_2023,chen2026finite}, underscoring both the generality of the phenomenon and its technological relevance, further highlighted by recent high AC-to-DC rectifier efficiencies in superconducting diodes~\cite{ingla_2025,castellani_2025}.

A particularly promising platform for realizing the SDE is the Rashba nanowire proximitized by a conventional $s$-wave superconductor and subjected to external magnetic fields, often referred to as a Majorana wire, where the interplay of spin–orbit coupling (SOC) and Zeeman fields naturally favors finite-momentum pairing and SDE~\cite{Legg_2022,bhowmik_2025,Picoli_2023}. In the minimal single-channel limit with linear SOC, however, the diode efficiency remains small, typically of the order of $\sim 2\%$, and requires the simultaneous presence of longitudinal and transverse magnetic-field components~\cite{Legg_2022}. Extensions including higher-order Rashba SOC relax some symmetry constraints, yet self-consistent calculations show that the achievable efficiencies remain limited~\cite{bhowmik_2025}. These studies predominantly address the strictly single-channel regime. In contrast, realistic nanowire devices generically host multiple occupied transverse subbands, where interchannel coupling and subband asymmetry can qualitatively modify superconducting correlations. A systematic investigation of the SDE in this experimentally relevant multichannel regime remains largely unexplored.

In this work, we address the superconducting diode effect in multichannel Rashba nanowires by explicitly considering both harmonic (cylindrical) and rectangular quantum-well confinement within a self-consistent Bogoliubov–de Gennes framework. We show that multichannel occupancy generically promotes asymmetric FF superconductivity, which serves as the microscopic origin of a strong nonreciprocal supercurrent response. The resulting current-driven FF state also stabilizes a topological superconducting phase supporting Majorana zero modes over an extended parameter regime, with the injected supercurrent providing direct control of the Cooper-pair momentum and the topological transition. We demonstrate that interchannel coupling qualitatively modifies the symmetry requirements for nonreciprocity and substantially enhances diode efficiency, reaching values of order $\sim 60\%$ depending on confinement geometry, with rectangular confinement further enabling a tunable sign reversal of the diode response. Our analysis of the pairing susceptibility reveals that a field-induced asymmetry in the pairing-susceptibility underlies the directional stabilization of Cooper pairing. These results establish multichannel Rashba nanowires as a realistic and versatile platform in which high-efficiency superconducting diode functionality and topological superconductivity naturally coexist.

The remainder of this article is organized as follows. In \sect{sec:mcn_harmonic}, we introduce the model Hamiltonian for a cylindrical Rashba nanowire with harmonic transverse confinement, proximitized by a three-dimensional ($3$D) $s$-wave superconductor, and describe the emergence of the FF superconducting state. In Sec.~\ref{subsec:topology_harmonic}, we examine the realization of topological superconductivity within the FF phase of the harmonically confined nanowire. The SDE arising from the interplay between Rashba spin–orbit coupling and external magnetic fields is analyzed in Sec.~\ref{subsec:sde_harmonic}, where we consider both interacting-channel and independent-channel limits. The Zeeman-field dependence of the diode efficiency is clarified through an analysis of the superconducting pairing susceptibility in \sect{subsec:susceptibility_harmonic}. We then turn to an alternative confinement geometry in \sect{sec:mcn_rectangular_well}, introducing the corresponding model Hamiltonian for a multichannel Rashba nanowire with rectangular quantum-well confinement, likewise proximitized by a $3$D $s$-wave superconductor. The topological properties of the FF state, the SDE, and the pairing susceptibility in this geometry are discussed in \sect{subsec:topology_rectangular_well}, \sect{subsec:sde_rectangular_well}, and \sect{subsec:susceptibility_rectangular_well}, respectively. Finally, we summarize our findings and present an outlook in \sect{sec:summ}.

\section{Multichannel Nanowire with Harmonic Confinement} \mylabel{sec:mcn_harmonic}

We consider a quasi one-dimensional (1D) Rashba nanowire, harmonically confined in the transverse ($y$, $z$) directions and free to propagate along $x$~\cite{Park_2017}. The nanowire of length $L$, placed adjacent to a three-dimensional $s$-wave superconductor, inherits superconducting correlations through the proximity effect. Under an externally applied magnetic field, the normal-state behavior of the system is governed by the 1D Rashba–Zeeman Hamiltonian

\beq
\mathcal{H}^{\rm HC}_{\rm NW} = \sum_{s,s^\prime} \int d\mathbf{r} \, \psi_s^\dagger(\mathbf{r}) h^{\rm HC}_{s s^\prime}(\mathbf{r})\psi_{s^\prime}(\mathbf{r}),\non
\eeq
\vspace{-0.5cm}
\beq
\!\!\hat{h}^{\rm HC}(\mathbf{r}) =  H^{\rm HC}_{0} (\mathbf{r}) + H_{\text{R}} (\mathbf{r}) + H_{\text{Z}} - \mu\,,
\label{eqn:nanowire_hc}
\eeq

\noin where $\mu$ is the chemical potential. The term $H^{\rm HC}_0$ represents the single-particle Hamiltonian of the quasi-1D nanowire, given by

\beq
H^{\rm HC}_0 (\mathbf{r}) = \frac{p_x^2 + p_y^2 + p_z^2}{2m} + U^{\rm HC}_c(y,z)\,.
\label{eq:kinetic_confinement}
\eeq

\noin Here, $m$ denotes the effective mass of the conduction electrons in the quasi-one-dimensional nanowire. The transverse confinement along the $y$ and $z$ directions is modeled by a harmonic potential, $U^{\rm HC}_c(y, z) = \tfrac{1}{2} m \omega_0^2 (y^2 + z^2)$, where $\omega_0$ is the characteristic angular frequency determining the confinement strength~\cite{Park_2017}. The corresponding effective diameter of the nanowire is estimated as $W = 2\sqrt{\hbar / (m \omega_0)}$. The contributions arising from Rashba spin-orbit coupling (SOC) ($H_{\mathrm{R}}$) and Zeeman interaction ($H_{\mathrm{Z}}$) are given by

\bea
H_{\text{R}} &=& -\frac{\alpha}{\hbar} p_x \sigma_y + \frac{\alpha}{\hbar} p_y \sigma_x\,, \\
H_{\text{Z}} &=& B_x \sigma_x + B_y \sigma_y\,,
\eea

\noin where $\alpha$ is the Rashba spin–orbit coupling strength, and $B_x$ and $B_y$ are the Zeeman energy terms corresponding to the magnetic field components along the $x$ and $y$ directions, respectively. The Pauli matrices $\sigma_{x,y,z}$ act on the spin degrees of freedom.

\begin{figure}[t]
    \centering
    \includegraphics[width=\linewidth]{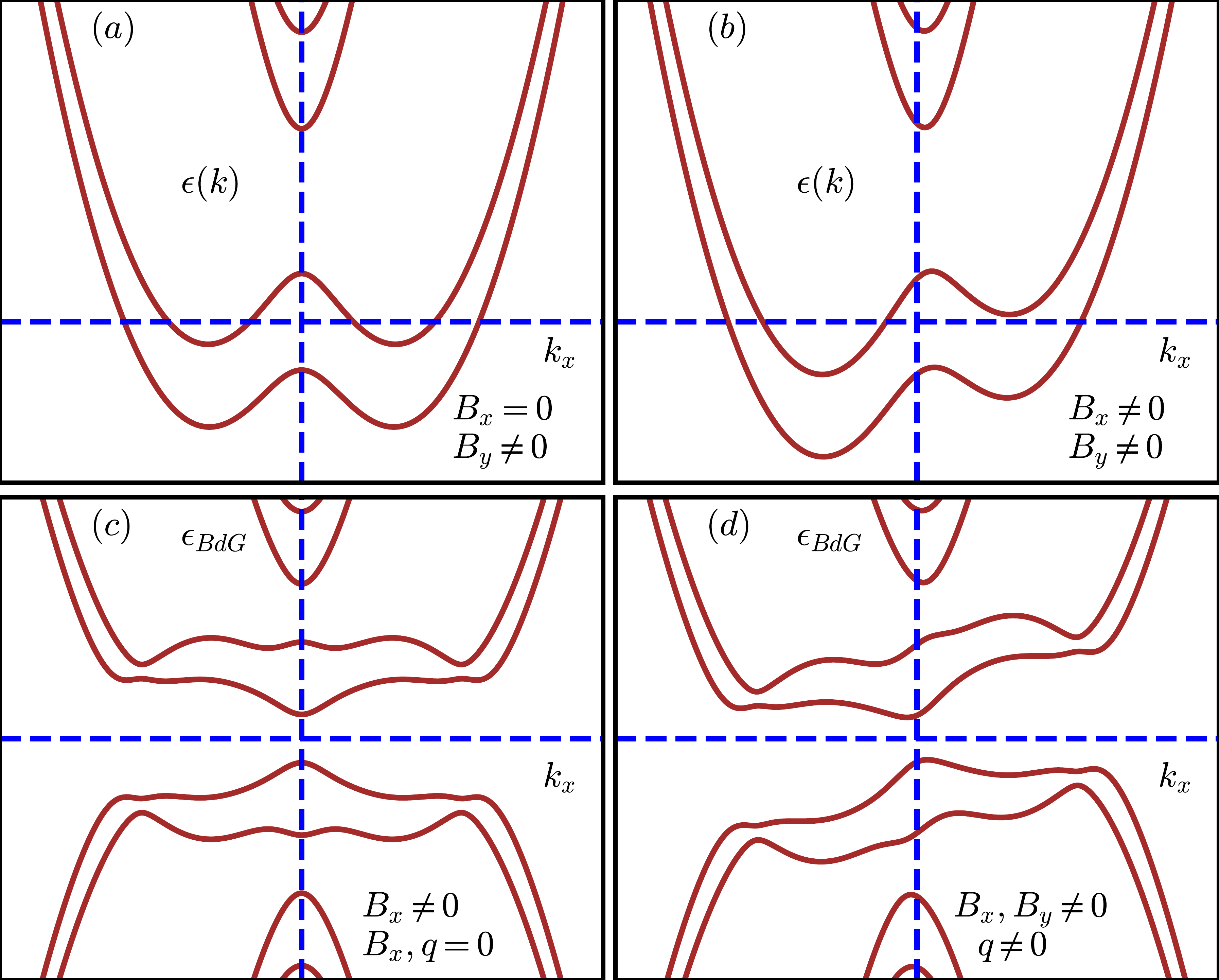}
    \caption{\textbf{Electronic and  Bogoliubov quasiparticle spectra of a multichannel nanowire under harmonic confinement::} Normal-state energy spectrum of the multichannel nanowire: (a) in the presence of an external magnetic field with $B_x \neq 0$ and $B_y = 0$, and (b) when both in-plane components are finite ($B_x \neq 0$, $B_y \neq 0$). Panels (c) and (d) show the corresponding Bogoliubov quasiparticle spectra. In (c), we consider $B_x \neq 0$, $B_y = 0$ with zero Cooper-pair momentum ($q = 0$), whereas in (d), both $B_x \neq 0$ and $B_y \neq 0$ are applied and a finite Cooper-pair momentum $q \neq 0$ is included.
}
    \label{spectrum}
\end{figure}

To simplify the analysis, we construct an effective one-dimensional (1D) Hamiltonian by integrating out the transverse ($y$, $z$) degrees of freedom. The combined contribution of the kinetic and confinement terms in Eq.~(\ref{eq:kinetic_confinement}), $(p_y^2 + p_z^2)/(2m) + U^{\rm HC}_c(y,z)$, leads to the following set of eigenvalues:

\beq
E^{\rm HC}_{n_y,n_z} = \hbar \omega_0 (n_y + n_z + 1) = \frac{4 \hbar^2}{m W^2} (n_y + n_z + 1)
\eeq

\noin where $n_y, n_z = 0, 1, 2, \dots$. The corresponding transverse eigenstates, denoted by $\phi^{\rm HC}_{n_y n_z s}(y, z)$ with spin index $s = \ua, \da$. The lowest three energy levels-associated with the eigenvalues $\hbar \omega_0$ and $2\hbar \omega_0$, are given by

\bea
\phi^{\rm HC}_{00s}(y,z) &=& \frac{2}{\sqrt{\pi} W} \, e^{-2(y^2 + z^2)/W^2} \, \chi_s\,, \\
\phi^{\rm HC}_{10s}(y,z) &=& \frac{4\sqrt{2} \, y}{\sqrt{\pi} W^2} \, e^{-2(y^2 + z^2)/W^2} \, \chi_s\,, \\
\phi^{\rm HC}_{01s}(y,z) &=& \frac{4\sqrt{2} \, z}{\sqrt{\pi} W^2} \, e^{-2(y^2 + z^2)/W^2} \, \chi_s\,.
\eea

\noin Here, $\chi_{\ua(\da)} = \tfrac{1}{\sqrt{2}}(1, \pm i)^T$ denote the eigenstates of the Pauli matrix $\sigma_y$. The transverse modes $\phi^{\rm HC}_{10s}(y, z)$ and $\phi^{\rm HC}_{01s}(y, z)$ are degenerate, each corresponding to an energy of $2\hbar \omega_0$. Notably, the mode $\phi^{\rm HC}_{01s}(y, z)$ does not couple to $\phi^{\rm HC}_{00s'}(y, z)$ through the spin–orbit interaction.

\beq
\int dy\,dz\, \phi_{00s'}^{\rm{HC}\dagger}(y,z) H_R \phi_{01s}^{\rm HC}(y,z) = 0
\eeq

\noin As a result, we restrict our attention to the lowest relevant subbands, $\phi^{\rm HC}_{00s}$ and $\phi^{\rm HC}_{10s}$, for analyzing the superconducting diode effect (SDE), since $\phi^{\rm HC}_{01s}(y, z)$ does not contribute to the modification of these lowest subbands. We then project the nanowire Hamiltonian in \eqn{eqn:nanowire_hc} onto the reduced subspace $\{ \phi^{\rm HC}_{00 \ua}, \phi^{\rm HC}_{00 \da}, \phi^{\rm HC}_{10 \ua}, \phi^{\rm HC}_{10 \da} \}$ and upon integrating out the transverse degrees of freedom, obtain the effective low-energy multichannel nanowire Hamiltonian, which is given by

\beq
\mathcal{H}_{\rm MCN}^{\rm HC} = \sum_{\substack{j,j^\prime \\s,s^\prime}} \int dx \, \psi_{js}^\dagger(x) h_{j j^\prime;s s^\prime}^{\prime} (x)\psi_{j^\prime s^\prime}(x),\non
\eeq
\vspace{-0.5cm}
\bea
\!\!\!\! h^{\prime} (x) &=&  H_{0}^{\prime} + H_{\text{R}}^{\prime} + H_{\text{Z}}^{\prime} - \mu\,. \non  \\
\text{\rm with,}\;\; H^{\prime}_0 &=& \frac{p_x^2}{2m} + E^{\rm HC}_+ + E^{\rm HC}_- \tau_z\,, \non \\
H^{\prime}_R &=& -\frac{\alpha}{\hbar}~p_x \tilde{\sigma}_z + \delta_{\rm HC}~\tilde{\sigma}_y \tau_y \non \,, \label{eq:Rashba_term}\\
H^{\prime}_Z &=& B_x \tilde{\sigma}_y + B_y \tilde{\sigma}_z\,. \label{eqn:mcn_nanowire_ham}
\eea

\noin The subscripts $js$ on the $\psi_{js}$ denote the transverse quantum numbers $j = 0,1$ and the spins $s = \ua, \da$. Here, $E^{\rm HC}_{\pm} = (E^{\rm HC}_{00} \pm E^{\rm HC}_{10})/2$. The Pauli matrices $\tilde{\sigma}_{x,y,z}$ and $\tau_{x,y,z}$ act in the spin space spanned by $\{\chi_{\ua}, \chi_{\da}\}$ and in the transverse-mode subspace, respectively. In \eqn{eq:Rashba_term}, the parameter $\delta_{\rm HC}$ quantifies the coupling between distinct transverse subbands with opposite spin orientations and is explicitly given by

\beq
\delta_{\rm HC} =\!\!\! \int dy \, dz \, \phi^{\rm{HC} \dagger}_{00\da}(y,z) \left(-i \alpha \frac{\partial}{\partial y} \sigma_x \right) \phi^{\rm HC}_{10\ua}(y,z) = \frac{\sqrt{2} \alpha}{W}
\eeq

\noin If $\delta_{\rm HC}=0$, the channels are non-interacting, the system reduces to a collection of independent single-channel Rashba nanowires — effectively a ``decoupled multichannel''. The Hamiltonian of multichannel nanowire in \eqn{eqn:mcn_nanowire_ham} can be written in momentum space

\beq
\mathcal{H}_{\rm MCN}^{\rm HC} = \sum_{k}\sum_{\substack{j,j^\prime \\s,s^\prime}}  c_{k,j,s}^\dagger h_{j j^\prime;s s^\prime}^{\prime} (k) c_{k,j^\prime, s^\prime},\non
\eeq
\vspace{-0.5cm}
\bea
h^{\prime} (k) &=& \frac{\hbar^2 k^2}{2m} + E^{\rm HC}_+ + E^{\rm HC}_- \tau_z -\alpha k \tilde{\sigma}_z + \delta_{\rm HC}~\tilde{\sigma}_y \tau_y \non \\ 
&+& B_x \tilde{\sigma}_y + B_y \tilde{\sigma}_z
\eea

\noin The multichannel nature of the nanowire gives rise to two additional spin-split Rashba bands compared to the single-channel case. While the external transverse Zeeman field $B_y$ induces an asymmetry in the nanowire spectrum, the longitudinal Zeeman field $B_x$ opens up a Zeeman gap, as shown in \fig{spectrum}(a,b).

We investigate the superconducting ground state of the Rashba nanowire in the presence of proximity-induced $s$-wave superconductor and externally applied supercurrents. Supercurrents can give rise to a superconducting state with finite-momentum Cooper pairs, known as the FF state. To capture this behavior, we introduce an attractive Hubbard-type interaction in the bulk $s$-wave superconductor,

\beq
\mathcal{H}_I = -\frac{U}{2}\sum{s,s^\prime} \int d^{3}{\bf r},
\psi_{s}^\dagger({\bf r})\psi_{s^\prime}^\dagger({\bf r})
\psi_{s^\prime}({\bf r})\psi_{s}({\bf r}), \label{eqn:hubbard_interaction}
\eeq

\noin where $U$ denotes the strength of the attractive interaction. We decouple $\mathcal{H}_I$ in the finite-momentum $s$-wave channel within the mean-field approximation~\cite{Daido_2022,akito_2022,Picoli_2023} and treat the resulting superconducting correlations as proximity-induced in the nanowire. This self-consistent mean-field framework effectively models an intrinsic one-dimensional superconductor with spin–orbit coupling. The induced correlations originate from the bulk superconductor and depend on factors such as interface transparency and magnetic fields. As a zeroth-order approximation, we assume a similar self-consistency condition as in the bulk, modulo a renormalized pairing gap—an assumption justified by the high symmetry of the parent superconductor, which suppresses fluctuation effects~\cite{Picoli_2023,bhowmik_2025}. Using the BdG mean-field formalism, the effective Hamiltonian of the nanowire then takes the form

\bea
\mathcal{H} & = & \frac{1}{2}\int\,dx\,\Psi^\dagger(x)\mathcal{H}_{BdG}(x)\Psi(x) + \mathcal{E}_0 \,,\non \\
\!\!\!\!\!\!\!\!\mathcal{H}_{BdG} (x) & = & 
\begin{bmatrix}
h'(x) & \hat{\Delta}(x) \\
-\hat{\Delta}^*(x) & -h'^*(x)
\end{bmatrix}; \, \hat{\Delta}(x)  =   -i\sigma_y \Delta  e^{iqx}\,,
\mylabel{eqn:BdG_Ham}
\eea

\begin{figure}[t]
    \centering
    \includegraphics[width=1\linewidth]{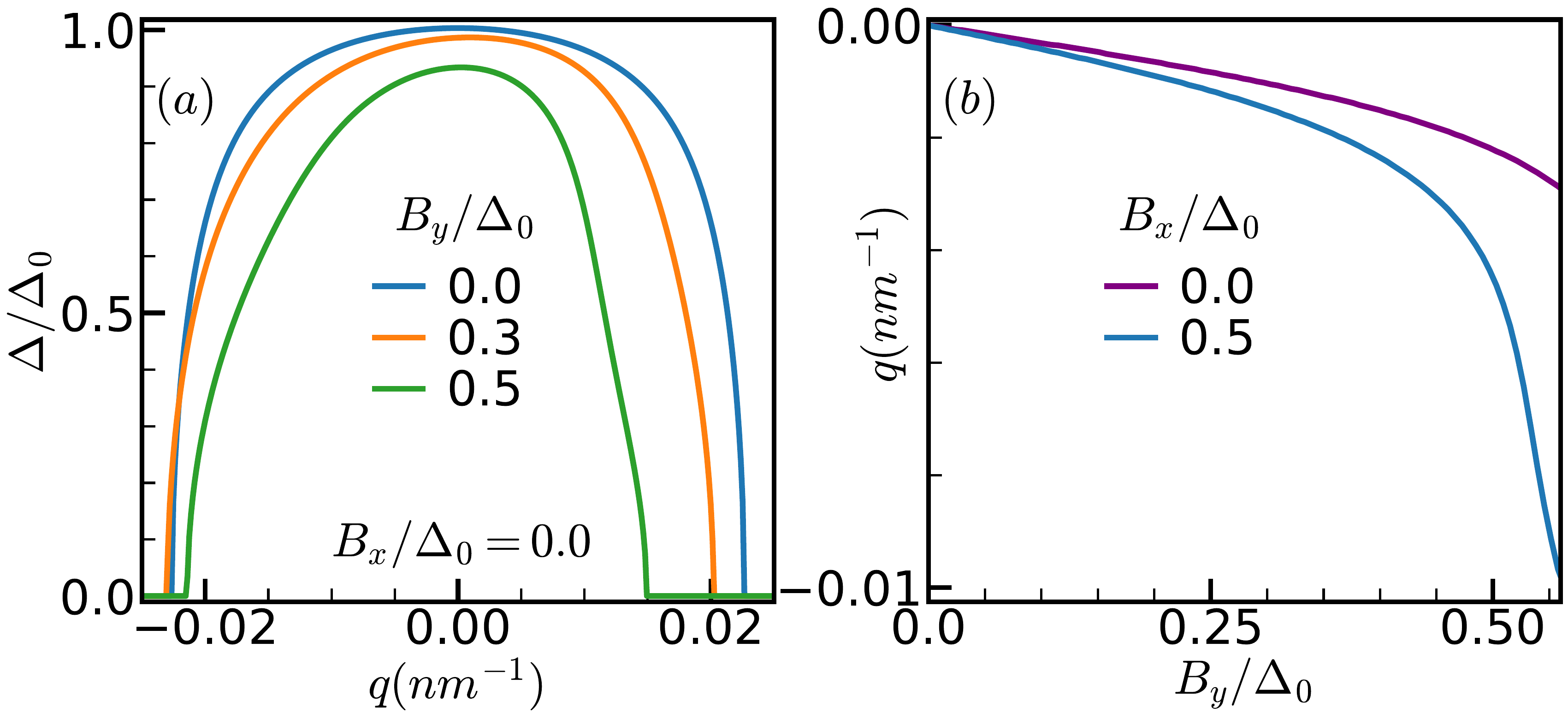} 
    \caption{\textbf{FF superconducting ground state in a harmonically confined nanowire:} (a) The self-consistent superconducting gap $\Delta(q)$ is shown as a function of the finite Cooper-pair momentum $q$ for a fixed $B_x = 0.0$ and varying $B_y$. Here, $\Delta_0$ denotes the superconducting order parameter in the absence of any external magnetic field.(b) The finite momentum value of the {FF} ground state, $q_0$, is plotted as a function of $B_y$. A stable {FF} ground state with $q_0 \neq 0$ emerges even when only one magnetic field component is present $(B_x = 0)$. The model parameters used are $(B_z, \mu, U, \beta^{-1}) = (0.0, 0, {13.35} \, \text{meV}, 0.1 \, \text{meV})$.}
    \label{fig:FFLO_state_hc}
\end{figure}

\noin where $\Psi(x) = [\psi(x), \psi^\dagger(x)]^T$, with $\psi(x) = [\psi_{0\uparrow}(x), \psi_{0\downarrow}(x), \psi_{1\uparrow}(x), \psi_{1\downarrow}(x)]^T$, denotes the Nambu spinor of the nanowire. The condensation energy is given by $\mathcal{E}_0 = (L/U)|\Delta|^2$, where $L$ is the nanowire length. The $s$-wave {FF} order parameter takes the form $\Delta(x) = \Delta (-i\sigma_y)e^{iqx}$, with $q$ representing the Cooper pair momentum. The Hamiltonian in \eqn{eqn:BdG_Ham} therefore describes the low-energy physics of the Rashba nanowire. To proceed, we express the mean-field Hamiltonian [\eqn{eqn:BdG_Ham}] in momentum space as

\bea
\mathcal{H} &=& \frac{1}{2} \sum_{k}\Psi_{k}^\dagger\mathcal{H}_{BdG} (k) \Psi_{k}+ \mathcal{E}_0 \,, \non\\
&=& \sum_{n,k} \left(E_{n,k} \gamma^\dagger_{n,k} \gamma_{n,k} - \frac{E_{n,k}}{2} \right) + \mathcal{E}_0 \,, \non 
\eea
\vspace{-0.3cm}
\beq
\text{\rm with,}\;\;\mathcal{H}_{BdG} (k) =
\begin{bmatrix}
h^\prime(k + \frac{q}{2}) & -i\sigma_y \Delta \\
i\sigma_y \Delta & - h^\prime(-k+\frac{q}{2})^*
\end{bmatrix}\,.
\eeq

\noin Here, $\Psi_k \equiv [C_{k+q/2}, C_{-k+q/2}^\dagger]^T$, where $C_{k+q/2} \equiv [c_{k+q/2,0,\ua}, c_{k+q/2,0,\da}, c_{k+q/2,1,\ua}, c_{k+q/2,1,\da}]^T$, and $E_{n,k}(q)$ denotes the Bogoliubov quasiparticle energies. The corresponding quasiparticle spectrum is illustrated in \fig{spectrum}(c,d), showing that the introduction of a finite pairing potential $\Delta$ opens a superconducting gap in the spectrum. Similar to the normal-state band structure, the transverse Zeeman field $B_y$ induces an asymmetry in the BdG spectrum, while the longitudinal field $B_x$ opens a Zeeman gap and plays a key role in driving the system into the topological phase.

The condensation energy $\Omega(q)$, defined as the difference in free energy per unit length between the superconducting and normal states~\cite{Kinnunen_2018,Daido_2022}, is given by

\beq
\Omega(q,\Delta) = F(q,\Delta) - F(q,0)\,,
\label{eqn:cond_energy}
\eeq

\noin where $F(q,\Delta)$ denotes the free-energy density of the nanowire,

\beq
F(q,\Delta) = -\frac{1}{L\beta}\sum_{n,k} \ln \left[1 + e^{-\beta E_{n,k}(q)}\right] + \frac{|\Delta(q)|^2}{U}\,,
\label{eqn:free_energy}
\eeq

\noin and $\beta = (k_B T)^{-1}$, with $k_B$ the Boltzmann constant and $T$ the temperature.

For a given $q$, the order parameter $\Delta(q)$ is determined self-consistently from the gap equation obtained by minimizing $F(q,\Delta)$ in \eqn{eqn:free_energy},

\beq
\Delta(q) = -\frac{U}{L} \sum_{n,k} \frac{\partial E_{n,k}}{\partial \Delta^*}\, n_F(E_{n,k})\,, \label{eqn:self_consistency}
\eeq

\noin where $n_F(E_{n,k}) = [1 + e^{\beta E_{n,k}}]^{-1}$ is the Fermi–Dirac distribution. Substituting the self-consistent $\Delta(q)$ into \eqn{eqn:cond_energy}, we minimize $\Omega(q)$ with respect to $q$ to obtain the equilibrium Cooper pair momentum $q_0$ corresponding to the true {FF} ground state. In this equilibrium configuration, the net charge current through the system vanishes. The zero-field BCS gap ($\mathbf{B}=0$) is denoted by $\Delta_0$, and the interaction strength $U$ is chosen such that the system lies in the weak-coupling BCS regime with $\Delta_0 \sim 1~\mathrm{meV}$. Unless otherwise stated, all results are presented in the low-temperature limit with $\beta^{-1} = 0.1~\mathrm{meV}$, well below $\Delta_0$.

The dependence of the self-consistent superconducting order parameter $\Delta(q)$ and the corresponding Cooper pair momentum $q_0$ in the FF ground state of the multichannel nanowire on the transverse Zeeman field $B_y$ is shown in \fig{fig:FFLO_state_hc}. As depicted in \fig{fig:FFLO_state_hc}(a), the superconducting gap collapses beyond critical momenta—above a positive threshold $q = q_c^+$ or below a negative threshold $q = q_c^-$. \fig{fig:FFLO_state_hc}(b) shows that $q_0$ increases linearly with $B_y$ in the weak-field regime ($B_y \ll \Delta_0$) for a fixed $B_x$. This linear dependence is key to the emergence of current-driven topology and the resulting superconducting diode effect (SDE), discussed in detail in the following sections.

\begin{figure}[t]
    \centering
    \includegraphics[width=\linewidth]{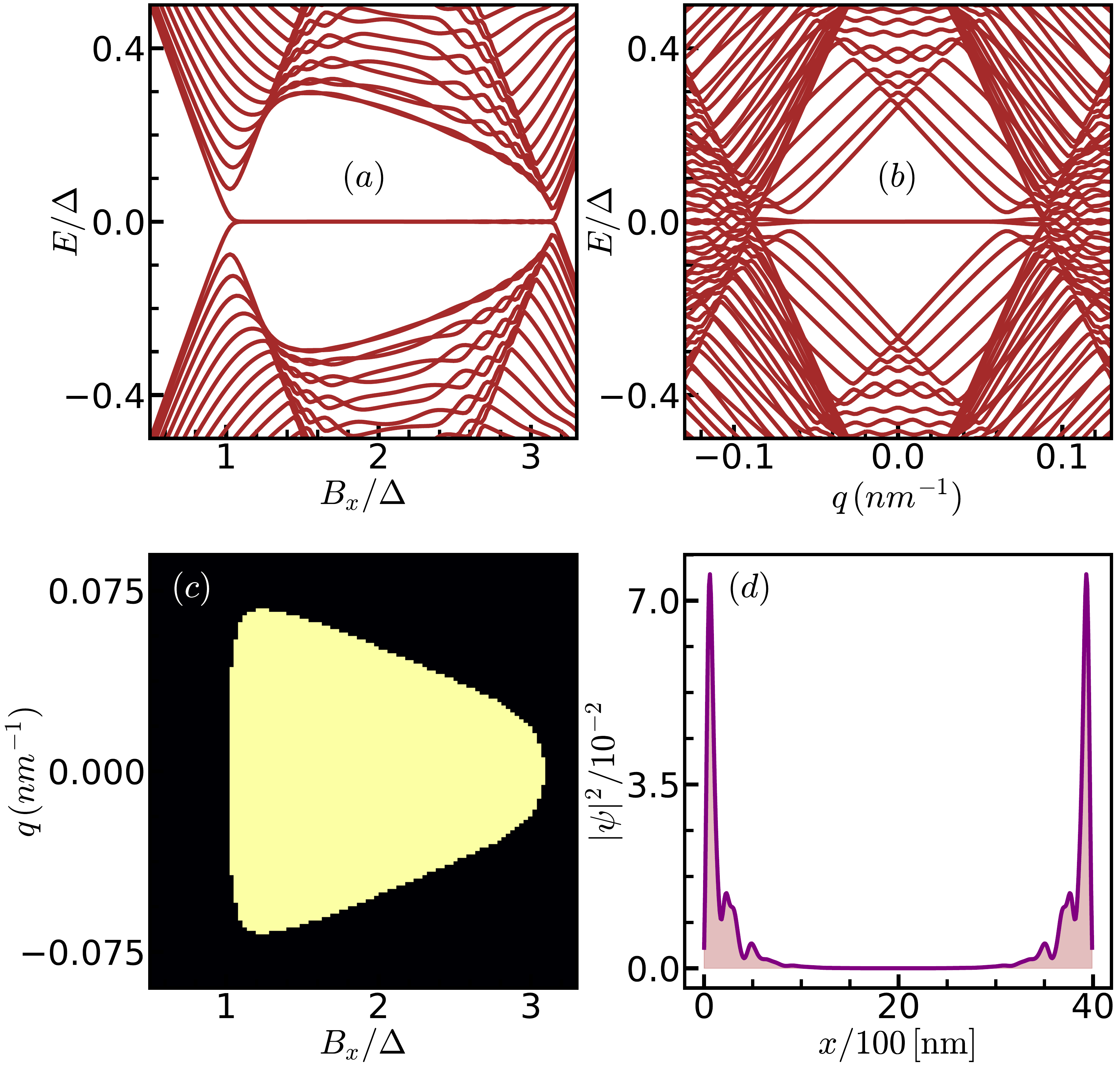}
    \caption{\textbf{Topological signatures of the FF superconducting state for harmonic confinement case:} (a) Low-energy spectrum as a function of the longitudinal magnetic field $B_x$, showing the emergence of zero-energy states associated with MZMs. (b) Evolution of the energy spectrum with Cooper-pair momentum for $B_x /\Delta_0 = 2$. (c) Phase diagram in the ($q, B_x$) plane, indicating the boundary between topologically trivial (black) and nontrivial (yellow) regions for a fixed pairing potential $\Delta = 0.5$. (d) Spatial probability density of a Majorana quasiparticle at $B_x / \Delta = 2$. The calculations were performed using $L = 4~\mu\text{m}$, $\Delta = 0.5~\text{meV}$, $\mu = 0.0~\text{meV}$, and $\alpha = 100~\text{meV·nm}$.}
    \label{fig:TSC}
\end{figure}

\subsection{Lattice Model}

To reveal the topological characteristics of the proposed setup and identify the signatures of Majorana zero modes (MZMs), we construct a minimal model on a one-dimensional lattice with open boundary conditions:

\begin{align}
\mathcal{H}_L &= \sum_{n=1}^{L} \Bigg[ \sum_{\substack{\tilde{\sigma},\tilde{\sigma}', \tau}} c^{\dagger}_{n, \tilde{\sigma}, \tau} \bigg\{ -t \delta_{\tilde{\sigma} \tilde{\sigma}'} - \frac{i\alpha}{2a} \left(\tilde{\sigma}_z\right)_{\tilde{\sigma}\tilde{\sigma}'}  \bigg\} c_{n+1, \tilde{\sigma}', \tau} \non\\  
&+ \sum_{\substack{\tilde{\sigma},\tilde{\sigma}' \\ \tau, \tau'}} c^{\dagger}_{n, \tilde{\sigma}, \tau} \bigg\{ \bigg((2t - \mu + E^{\rm HC}_{+})\delta_{\tau \tau'} + E^{\rm HC}_{-} (\tau_z)_{\tau \tau'}\bigg) \delta_{\tilde{\sigma} \tilde{\sigma}'} \non \\
&+ \bigg( B_x (\tilde{\sigma_y})_{\tilde{\sigma} \tilde{\sigma}'} + B_y (\tilde{\sigma_z})_{\tilde{\sigma} \tilde{\sigma}'} \bigg) \delta_{\tau \tau'} + \delta (\tilde{\sigma_y})_{\tilde{\sigma} \tilde{\sigma}'} (\tau_y)_{\tau \tau'} \bigg\} c_{n, \tilde{\sigma}', \tau'} \non \\
&+ \sum_{\tau} \Delta e^{iqna} \bigg(c^{\dagger}_{n, \ua, \tau} c^{\dagger}_{n, \da, \tau} \bigg) + \text{\rm h.c.} \Bigg] \,.\label{eqn:real_space_ham}
\end{align}

\begin{figure*}[t]
    \centering
    \includegraphics[width=\textwidth]{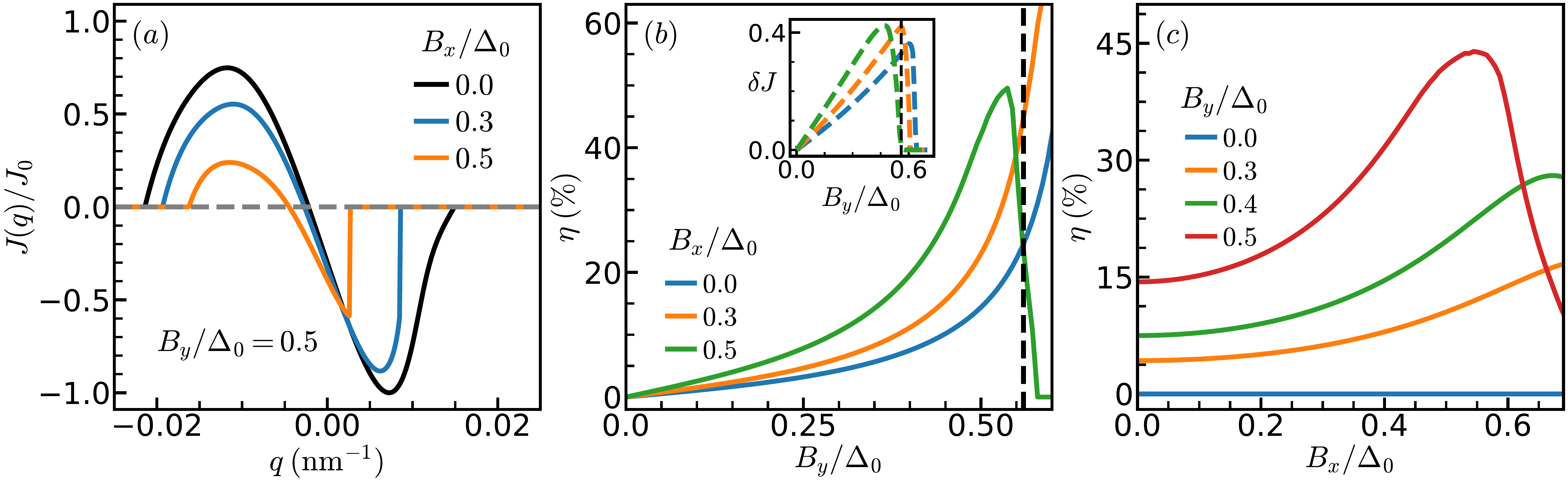}
    \caption{\textbf{Supercurrent density and diode efficiency in a harmonically confined nanowire with interacting channels ($\delta_{\rm HC} \neq 0$):}(a) Supercurrent density $J(q)$ as a function of the Cooper-pair momentum $q$ for fixed $B_y / \Delta = 0.5$ and different values of $B_x$. A clear asymmetry appears between the positive and negative critical supercurrents, and importantly, nonreciprocity persists even when only the transverse magnetic field is present ($B_y \neq 0$, $B_x = 0$). (b,c) The corresponding diode efficiency $\eta$ as a function of $B_y / \Delta$ for several fixed values of $B_x$ (panel b) and as a function of $B_x / \Delta$ for several fixed values of $B_y$ (panel c). The model parameters are $(B_z, \alpha, \mu, U, \beta^{-1}) = (0.0, 100~\text{meV - nm}, 0.0, {13.35}~\text{meV}, 0.1~\text{meV})$.}

    \label{fig:efficiency_interacting}
\end{figure*}

\noin The operator $c_{n,\tilde{\sigma},\tau}$ ($c_{n,\tilde{\sigma},\tau}^\dagger$) annihilates (creates) a fermion with spin $\tilde{\sigma}$ at lattice site $n$ in subband $\tau$. The chemical potential is denoted by $\mu$, and the nearest-neighbor hopping amplitude is $t = \hbar^2 / (2ma^2)$, where $m$ is the effective mass and $a$ is the lattice constant. The Zeeman fields induced by external magnetic fields applied along the $x$ and $y$ directions are represented by $B_x$ and $B_y$, respectively. The total length of the nanowire is $L = Na$, with $N$ being the number of lattice sites.

\subsection{Topologically protected MZMs in the FF superconducting phase}\mylabel{subsec:topology_harmonic}

To demonstrate the emergence of topological superconductivity in the multichannel Rashba nanowire, we analyze the quasiparticle spectrum obtained from the lattice Hamiltonian in \eqn{eqn:real_space_ham} under open boundary conditions, as shown in \fig{fig:TSC}. \fig{fig:TSC}(a),(b) present the evolution of the low-energy spectrum as functions of the longitudinal Zeeman field $B_x$ and the {FF} momentum $q$. For small $B_x$, the system remains in a trivial superconducting phase characterized by a fully gapped spectrum. As $B_x$ increases, the bulk gap closes and subsequently reopens beyond a critical field, marking a topological phase transition. In this regime, zero-energy modes appear and remain pinned near $E = 0$ over a finite range of $B_x$, as shown in \fig{fig:TSC}(a). The panel \fig{fig:TSC}(b) further highlights the emergence of current-driven topology, where the interplay between finite-momentum Cooper pairing ($q \neq 0$) and the Zeeman field drives the system into a topological superconducting phase. Importantly, the topological phase can now be tuned by an external control parameter—namely, the applied supercurrent that sets the finite pairing momentum $q$—offering a practical route to manipulate topological superconductivity without relying solely on magnetic fields. The parameter space supporting Majorana zero modes (MZMs) is mapped in \fig{fig:TSC}(c), which shows that MZMs persist over a broad range of $B_x$ and $q$, underscoring the robustness of the topological phase. The real-space probability density in \fig{fig:TSC}(d) confirms the end-localized nature of these zero-energy states, characteristic of MZMs, although their spatial confinement is weaker than in the single-channel case due to inter-subband hybridization. The coexistence of a reopened bulk gap, finite-momentum pairing, and current-tunable zero modes thus establishes the realization of a current-driven topological superconducting phase in the {FF} multichannel Rashba nanowire.

\begin{figure}[b]
    \centering
    \includegraphics[width=1\linewidth]{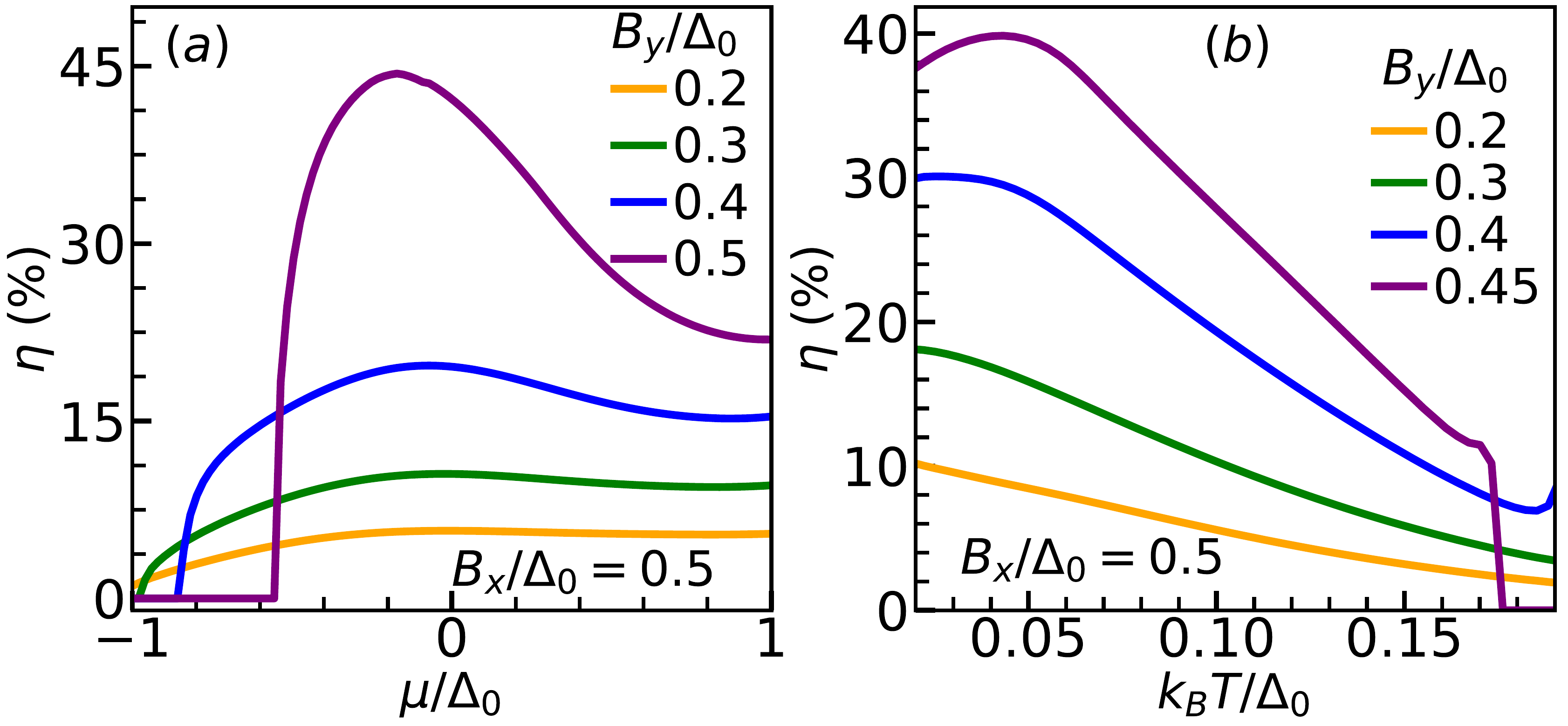}
    \caption{\textbf{Effect of chemical potential and temperature on superconducting diode efficiency of a harmonically confined nanowire:} (a) The plot shows how the superconducting diode efficiency varies with the chemical potential $\mu$ for a fixed $B_x/\Delta_0 = {0.5}$ and different values of $B_y$. (b) The second panel illustrates how the efficiency changes with temperature for a fixed $B_x/\Delta_0 ={0.5}$ while varying \(B_y\). The calculations are performed using following model parameters $(B_z,\alpha,  \mu, U) = (0.0, 100~\mathrm{meV-nm},  0.0, {10.35}~\mathrm{meV})$.}
    \label{fig:eff_T_mu}
\end{figure}

\begin{figure*}[t]
    \centering
    \includegraphics[width=\textwidth]{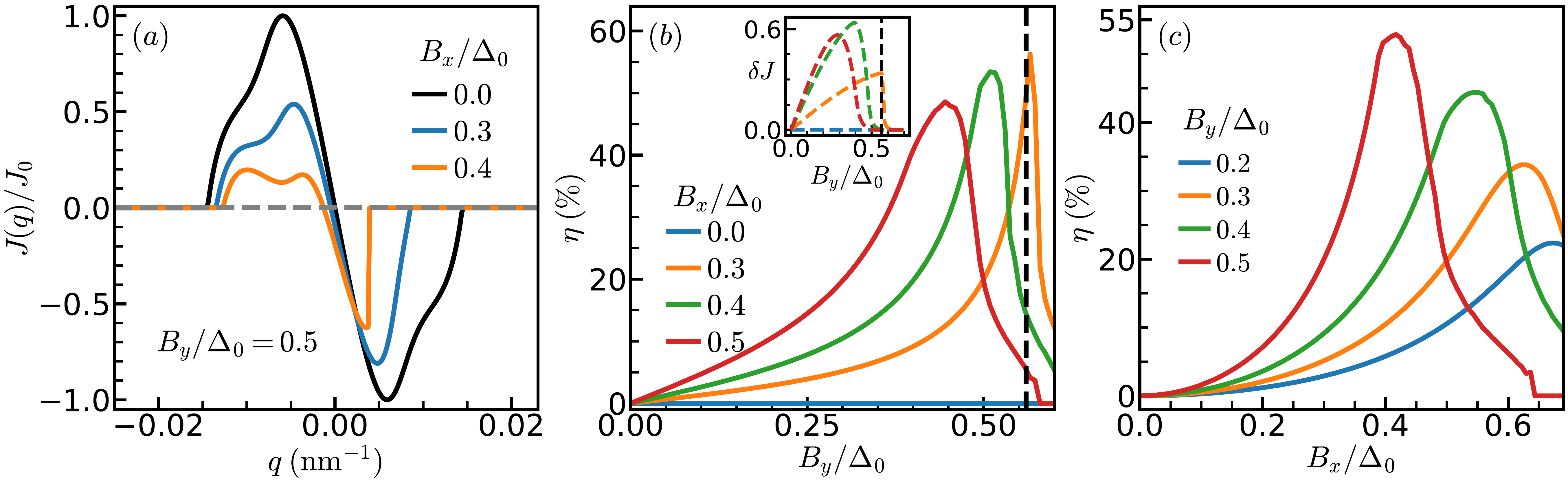}
   \caption{\textbf{Supercurrent density and diode efficiency in a harmonically confined nanowire with independent channels ($\delta_{\rm HC} = 0$):} (a) Supercurrent density $J(q)$ as a function of the Cooper-pair momentum $q$ for fixed $B_y / \Delta = 0.5$ and different values of $B_x$. A clear asymmetry between positive and negative critical currents requires both magnetic-field components to be nonzero ($B_y \neq 0$, $B_x \neq 0$). (b,c) Diode efficiency $\eta$ as a function of $B_y / \Delta$ for fixed values of $B_x$ (b) and as a function of $B_x / \Delta$ for fixed values of $B_y$ (c). Model parameters: $(B_z, \alpha, \mu, U, \beta^{-1}) = (0.0, 100~\mathrm{meV\!-\!nm}, 0.87, 12.63~\mathrm{meV}, 0.1~\mathrm{meV})$.}

   \label{fig:zero_band_mixing}
\end{figure*}

\subsection{Nonreciprocal Superconducting Transport}\mylabel{subsec:sde_harmonic}

The supercurrent density $J(q)$ is obtained from the condensation energy $\Omega(q,\Delta)$ [\eqn{eqn:cond_energy}] as~\cite{Yuan_2022,Daido_2022,Legg_2022}

\beq
J(q) = -2e\frac{\partial \Omega(q,\Delta)}{\partial q}.\label{SDEcurrent}
\eeq

\noin The sign of $J(q)$ determines the direction of the supercurrent flow. The critical supercurrents along and opposite to the flow direction, within the stable superconducting range $q_{c}^- \leq q \leq q_{c}^+$, are denoted by $J_{c}^+$ and $J_{c}^-$, respectively. A finite SDE arises when $|J_{c}^+| \neq |J_{c}^-|$, and its efficiency (or quality factor) is defined as~\cite{Yuan_2022}

\beq
\eta = \frac{|J_{c}^+| - |J_{c}^-|}{|J_{c}^+| + |J_{c}^-|}.\label{eqn:SDEefficiency}
\eeq

\noin To evaluate $\eta$, we first compute $J(q)$ over the range $q_{c}^- \leq q \leq q_{c}^+$ using the corresponding self-consistent solutions $\Delta(q)$ for each $q$ (as illustrated in \fig{fig:FFLO_state_hc}(a). The critical currents $J_{c}^+$ and $J_{c}^-$ are then extracted from the extrema of $J(q)$, allowing a numerical evaluation of the diode efficiency. We then analyze the SDE in the Rashba nanowire under two distinct scenarios: (i) interacting channels and (ii) independent channels, considering a range of relevant system parameters as discussed in the following subsections.

\subsubsection{Case-I: Interacting channels}

We analyze the SDE in the interacting-channel regime $(\delta_{HC} \neq 0)$ by studying how the supercurrent density $J(q)$ and the diode efficiency $\eta$ respond to variations in key system parameters, as presented in \fig{fig:efficiency_interacting}. As shown in \fig{fig:efficiency_interacting}(a), $J(q)$ exhibits a clear asymmetry about $q = 0$ even in the absence of a longitudinal Zeeman field ($B_x = 0$) when a finite transverse component ($B_y \neq 0$) is applied. This behavior contrasts with that of a single-channel nanowire hosting only linear SOC~\cite{Legg_2022,bhowmik_2025}. When $B_x$ is introduced and gradually increased while keeping $B_y$ finite, the asymmetry in $J(q)$ becomes more pronounced, giving rise to unequal critical currents ($|J_{c}^+| \neq |J_{c}^-|$) and hence a finite SDE. The emergence of this nonreciprocity can be traced to the interplay between the finite-momentum {FF} pairing and the imbalance in the critical momenta ($|q_{c}^+| \neq |q_{c}^-|$), which manifests when a transverse Zeeman field is present, as seen in \fig{fig:efficiency_interacting}(a).

\fig{fig:efficiency_interacting}(b) and \fig{fig:efficiency_interacting}(c) display the behavior of the diode efficiency $\eta$ as a function of $B_y$ for fixed $B_x$ and as a function of $B_x$ for fixed $B_y$, respectively. Notably, a finite efficiency persists even for $B_x = 0$ when $B_y \neq 0$, highlighting a distinct multichannel effect absent in single-channel Rashba nanowires with linear SOC~\cite{Legg_2022,bhowmik_2025}. In both parameter sweeps, $\eta$ grows nearly linearly with increasing field strength at low values, reaches a pronounced maximum at intermediate fields, and subsequently decreases as the Zeeman fields become stronger. This nonmonotonic behavior originates from the interplay between two competing effects: the enhancement of band-structure asymmetry induced by the magnetic fields and the simultaneous suppression of superconducting pairing susceptibility as the Zeeman energy increases~\cite{Yuan_2022,Turini2022}. At sufficiently large $B_x$ and $B_y$, the self-consistent superconducting gap collapses, driving $\eta$ to zero; however, the Bogoliubov quasiparticle spectrum can remain gapped, resembling that of a trivial insulating phase. The underlying mechanisms responsible for this competition are discussed in more detail below. As shown in \fig{fig:efficiency_interacting}(b), the diode efficiency can reach values of up to $\sim 60\%$. However, such large efficiencies are not expected to be experimentally relevant, since the asymmetry in the critical currents decreases beyond $B_y/\Delta_0 \approx 0.55$, as indicated by the inset of \fig{fig:efficiency_interacting}(b). In this regime, the critical currents become very small, leading to an apparent enhancement of $\eta$ due to the reduced denominator in its definition, rather than a genuinely strong nonreciprocal response.

The dependence of $\eta$ on the chemical potential $\mu$ and temperature $T$ is summarized in \fig{fig:eff_T_mu}. In \fig{fig:eff_T_mu}(a), $\eta$ is shown as a function of $\mu$ for several values of $B_y$ at a fixed ${B_x/\Delta_0 = 0.5}$. The efficiency exhibits a pronounced nonmonotonic dependence on $\mu$, reaching its maximum near moderate carrier densities where multiple subbands contribute to transport. For both low and high $\mu$, $\eta$ decreases as the system approaches either the single-channel or deep metallic limits, emphasizing the sensitivity of the SDE to band filling and the number of active conduction channels. The temperature dependence of $\eta$, presented in \fig{fig:eff_T_mu}(b), shows a gradual suppression with increasing $T$ for all $B_y$. The efficiency remains nearly constant at low temperatures, begins to decline around $k_B T \sim 0.1/\Delta_0$, and eventually vanishes near the critical temperature where the superconducting order parameter collapses. These results demonstrate that optimal diode performance occurs at intermediate $\mu$ and low $T$, where multichannel coherence and finite-momentum pairing are strongest, providing ideal conditions for realizing robust nonreciprocal superconductivity in the interacting-channel regime.

\subsubsection{Case-II: Independent channels}

We next investigate the SDE in the limit where the transverse channels are independent, i.e., when inter-channel coupling is absent $(\delta_{HC}=0)$. The behavior of the supercurrent density $J(q)$ and the corresponding diode efficiency $\eta$ for this case is shown in \fig{fig:zero_band_mixing}. Unlike the interacting-channel scenario, $J(q)$ remains symmetric about $q = 0$ when only one of the Zeeman field components is finite. Consequently, a finite $\eta$ appears only when both magnetic fields, $B_x$ and $B_y$, are simultaneously nonzero, consistent with previous studies on single-channel Rashba nanowires with linear SOC~\cite{Yuan_2022,Legg_2022,bhowmik_2025}. As illustrated in \fig{fig:zero_band_mixing}(a), increasing either $B_x$ or $B_y$ enhances the asymmetry in $J(q)$, leading to nonreciprocal critical currents, $|J_{c}^+| \neq |J_{c}^-|$, and thus a finite diode response. The field dependence of $\eta$, presented in \fig{fig:zero_band_mixing}(b) and \fig{fig:zero_band_mixing}(c), follows a similar nonmonotonic trend as in the interacting case—growing linearly at low fields, reaching a pronounced maximum at intermediate strengths, and diminishing as superconductivity weakens at high fields. Notably, in the independent-channel limit the maximum diode efficiency reaches approximately $50\%$, which is comparable to the strictly single-channel case with linear spin--orbit coupling. In contrast, when inter-channel coupling is present, a finite diode efficiency of about $20\%$ already appears even at $B_x = 0$. In the presence of both magnetic fields, the maximum efficiency is further enhanced and becomes comparable to that of the single-channel system. As indicated by  \fig{fig:zero_band_mixing}(b), the diode efficiency in the independent-channel limit can reach values as high as $\sim 55\%$. Similar to the interacting case, this large efficiency occurs in a regime where the critical currents are strongly suppressed, leading to an apparent enhancement of $\eta$ rather than a robust increase in current nonreciprocity [see the inset of \fig{fig:zero_band_mixing}(b)].

\subsection{Superconducting Pairing susceptibility} \mylabel{subsec:susceptibility_harmonic}

\begin{figure}[t]
    \centering
    \includegraphics[width=1\linewidth]{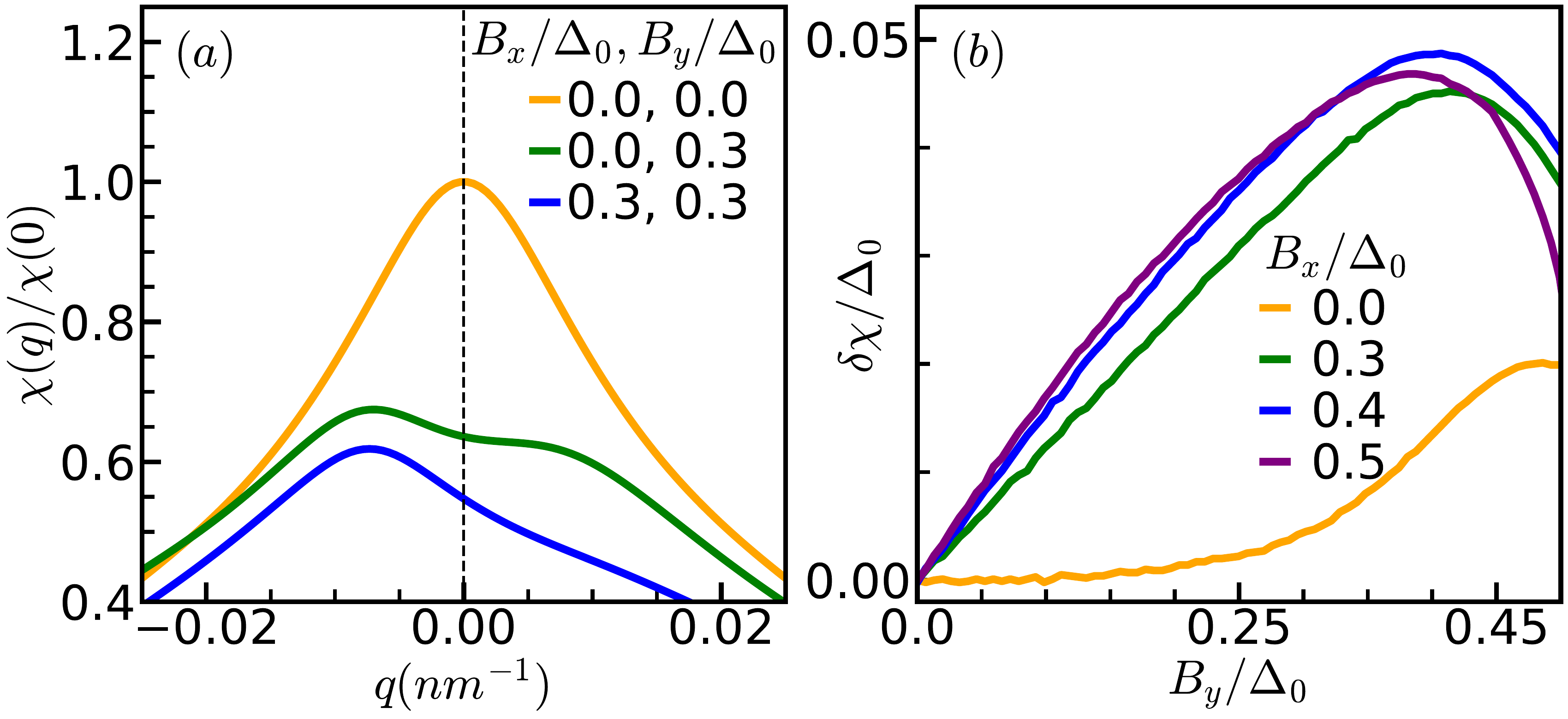}
    \caption{\textbf{Superconducting Pairing Susceptibility of a harmonically confined nanowire:} (a) Superconducting pairing susceptibility as a function of the finite Cooper-pair momentum $q$ for different values of the magnetic fields $B_x$ and $B_y$. (b) Difference in the pairing susceptibility evaluated at the Cooper-pair momenta ($q_c$), where the supercurrent is maximized, plotted as a function of the transverse magnetic field $B_y$ for several values of $B_x$.}
    \label{Susceptibility}
\end{figure}

To understand how the SDE efficiency depends on the Zeeman fields, we calculate the superconducting pairing susceptibility $\chi(q)$. This quantity characterizes how easily electrons with momenta $k+q/2$ and $-k+q/2$ form Cooper pairs carrying a finite total momentum $q$. It is given by~\cite{yuan_2021}

\bea
\chi(q) = k_{B}T \sum_{k, i\omega_{n}} \operatorname{Tr} && \bigg[\mathcal{J}^{\dagger}(\hat{k})\mathcal{G}\big(k+\frac{q}{2}, i\omega_{n}\big) \non \\
&& \mathcal{J}(\hat{k}) \mathcal{G}^{T}\big(-k+\frac{q}{2}, -i\omega_{n}\big)\bigg]\,,
\eea

\noin where $\mathcal{G}(k, i\omega_{n}) = [i\omega_{n} - H(k)]^{-1}$ is the single-particle Green’s function, $\omega_{n}=(2n+1)\pi k_{B}T$ denotes the fermionic Matsubara frequency at temperature $T$, and $\mathcal{J}(\hat{k})$ represents the superconducting form factor of the Cooper pair.

To evaluate the trace over internal degrees of freedom, we expand the Green’s function in the eigenbasis of the normal-state Hamiltonian as
$\mathcal{G}(k,i\omega_n)=\sum_{m}|m,k\rangle\langle m,k|/(i\omega_n-\epsilon_m)$.
This yields the pairing matrix elements
$M_{\lambda m}(k,q)=\langle \lambda,k+q/2|\mathcal{J}|m,-k+q/2\rangle$,
where $\mathcal{J}=(\tau_0 \otimes i\sigma_y)\mathcal{K}$ for $s$-wave pairing, $\mathcal{K}$ denotes complex conjugation, and $\lambda,m$ label the normal-state bands. The susceptibility can then be written in the band-resolved form

\beq
\chi(q)=k_{B}T\sum_{k}\sum_{\lambda,m,i\omega_n}
\frac{|M_{\lambda m}(k,q)|^2}{(i\omega_n-\epsilon_\lambda)(-i\omega_n-\epsilon_m)}\,.
\eeq

\noin The Matsubara frequency sum can be carried out analytically using contour integration, yielding

\beq
k_{B}T\sum_{i\omega_n}
\frac{1}{(i\omega_n-\epsilon_\lambda)(-i\omega_n-\epsilon_m)}
=\frac{1-f(\epsilon_\lambda)-f(\epsilon_m)}{\epsilon_\lambda+\epsilon_m}\,,
\eeq

\noin where $f(\epsilon)=[e^{\epsilon/k_{B}T}+1]^{-1}$ is the Fermi–Dirac distribution. Consequently, the static pairing susceptibility takes the final form

\beq
\chi(q)=\sum_{k}\sum_{\lambda,m}
\frac{1-f(\epsilon_\lambda)-f(\epsilon_m)}
{\epsilon_\lambda(k+q/2)+\epsilon_m(-k+q/2)}
|M_{\lambda m}(k,q)|^2\,.
\label{eqn:Susceptibility}
\eeq

When the pairing susceptibility becomes asymmetric, i.e., $\chi(q)\neq\chi(-q)$, it indicates a directional preference for finite-momentum Cooper pairing. This asymmetry directly translates into nonreciprocal supercurrent transport and thus provides a microscopic origin for the superconducting diode effect.

In Fig.~\ref{Susceptibility}(a), we show the superconducting pairing susceptibility as a function of the finite Cooper-pair momentum $q$. When a transverse magnetic field $B_y$ is applied, the susceptibility profile becomes clearly asymmetric, meaning $\chi(q)\neq\chi(-q)$. This appears as unequal peak heights at $+q$ and $-q$, indicating that Cooper pairing is energetically more favorable in one momentum direction over to the other. In simple terms, the system prefers Cooper pairs moving in a particular direction.

In Fig.~\ref{Susceptibility}(b), we quantify the pairing asymmetry by evaluating the difference between the pairing susceptibilities at the critical Cooper-pair momenta $q$ corresponding to the maximum supercurrent. We observe that the pairing asymmetry increases with increasing $B_y$, indicating an enhancement of the directional superconducting response. However, beyond a certain magnetic-field strength, the pairing asymmetry begins to decrease, resulting in a nonmonotonic dependence at strong fields.

\section{Multichannel Nanowire with Rectangular Quantum Well Confinement} \mylabel{sec:mcn_rectangular_well}

To assess the robustness and generality of our results with respect to the confinement geometry, we now consider a multichannel Rashba nanowire subject to a rectangular quantum well potential in the transverse directions. This alternative confinement geometry enables us to analyze how the modified subband structure and inter-channel coupling affect finite-momentum pairing, nonreciprocal transport, and topological properties beyond the harmonic approximation. We consider a rectangular quantum well with dimensions $L_x$, $L_y$, and $L_z$, placed in close proximity to a three-dimensional bulk $s$-wave superconductor. Assuming strong confinement along the $\hat{z}$ direction such that $L_z \ll L_y, L_x$, only the lowest subband corresponding to the $\hat{z}$-axis eigenstate is occupied. The single-particle hamiltonian then takes the usual form of a two-dimensional semiconducting nanowire in the presence of spin–orbit interaction and an external magnetic field,  

\beq
\mathcal{H}^{\rm RW}_{\rm NW} = \sum_{s,s^\prime} \int d\mathbf{r} \,
\psi_s^\dagger(\mathbf{r})\, h^{\rm RW}_{s s^\prime}(\mathbf{r})\,\psi_{s^\prime}(\mathbf{r}) \non\,, 
\eeq

\vspace{-0.5cm}

\beq
\hat{h}^{\rm RW}_{ss'}(\mathbf{r}) =  H^{\rm RW}_{0}(\mathbf{r}) + H_{\text{R}}(\mathbf{r}) + H_{\text{Z}} - \mu\,.
\label{eqn:nanowire_rw}
\eeq

\noin In the present geometry, the Hamiltonian $H^{\rm HC}_0(\mathbf{r})$ of the harmonically confined cylindrical nanowire is replaced by $H^{\rm RW}_0(\mathbf{r})$, describing a nanowire subject to rectangular well confinement.

\beq
H^{\rm RW}_{0}(\mathbf{r}) = \frac{p_x^{2} + p_y^{2}}{2m} + U^{\rm RW}_c(y)\,,
\label{eqn:2d_kinetic}
\eeq

\noin where $U^{\rm RW}_c(y)$ denotes the hard-wall confinement potential along the transverse $\hat{y}$ direction~\cite{Lutchyn2011}.

\beq
U^{RW}_c(y)=
\begin{cases}
0, & 0< y < L_y,\\
\infty, & \text{otherwise}\,.
\end{cases}
\eeq

\noin To construct an effective one-dimensional multichannel nanowire Hamiltonian, analogous to the harmonic confinement case discussed earlier, we first diagonalize the transverse Hamiltonian $p_y^2/2m + U^{\rm RW}_c(y)$. This yields a discrete set of transverse subband energies $E^{\rm RW}_n = (\hbar^2 \pi^2 n^2)/(2mL_{y}^2)$ together with the corresponding transverse eigenmodes

\beq
\phi^{\rm RW}_{ns}(y) = \sqrt{\frac{2}{L_y}}\, \sin\!\left(\frac{n\pi y}{L_y}\right)\chi_s\,,
\eeq

\noin Here, $n=1,2,\dots$ labels the transverse subbands and $\chi_{\uparrow(\downarrow)} = (1/\sqrt{2})(1,\pm i)^T$ denote the eigenstates of $\sigma_y$. Unlike the cylindrical nanowire with harmonic confinement, the rectangular quantum well does not support orbital degeneracies between higher transverse modes, apart from the trivial spin degeneracy. Projecting the two-dimensional (2D) Hamiltonian in \eqn{eqn:nanowire_rw} onto the low-energy subspace spanned by the lowest two transverse subbands, $\{ \phi^{\rm RW}_{1\ua}, \phi^{\rm RW}_{1\da}, \phi^{\rm RW}_{2\ua}, \phi^{\rm RW}_{2\da} \}$, and integrating over the transverse coordinate $y$, we arrive at an effective 1D multichannel nanowire Hamiltonian,

\begin{figure}[t]
    \centering
    \includegraphics[width=1\linewidth]{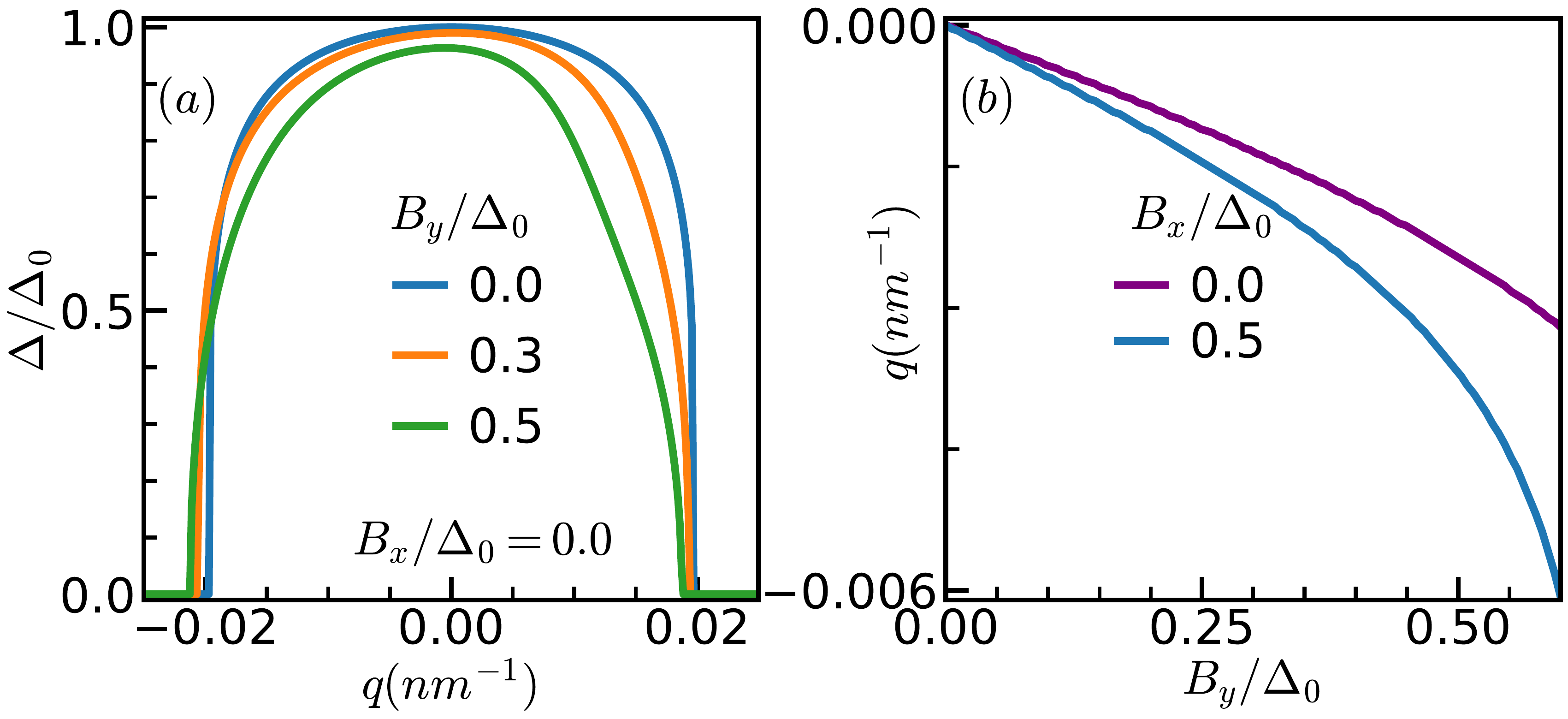}     
    \caption{\textbf{FF superconducting ground state in a nanowire with rectangular well confinement:} (a) The self-consistent superconducting order parameter $\Delta(q)$ is plotted as a function of the finite Cooper-pair momentum $q$ for a fixed longitudinal magnetic field $B_x = 0$ and different values of the transverse field $B_y$. The quantity $\Delta_0$ denotes the superconducting gap in the absence of any magnetic field. In the presence of a finite magnetic field $B_y\neq0$, the critical momenta become asymmetric, i.e., $q_c^{+} \neq q_c^{-}$. (b) The optimal finite momentum of the {FF} ground state, $q_0$, is shown as a function of $B_y$. A stable {FF} phase with $q_0 \neq 0$ emerges even when only a single magnetic-field component is applied ($B_x = 0$). The calculations are performed using the parameters $(B_z, \mu/\Delta_0, U, \beta^{-1}) = (0.0, 0, 15.47~\text{meV}, 0.1~\text{meV})$.}
    \label{fig:FFLO_state_rw}
\end{figure}

\beq
\mathcal{H}^{\rm RW}_{\rm MCN} = \sum_{\substack{j,j^\prime \\ s,s^\prime}} \int dx \, \psi_{js}^\dagger(x)\, \tilde{h}_{j j^\prime; s s^\prime}(x)\, \psi_{j^\prime s^\prime}(x)\,, \non
\eeq

\vspace{-0.5cm}

\bea
\tilde{h} (x) &=& \tilde{H}_{0} + \tilde{H}_{\text{R}} + \tilde{H}_{\text{Z}} - \mu\,, \non\\
\!\!\!\!\!\!\!\!\!\!\!\!\!\!\!\!\!\!\!\!\!\!\!\! \text{\rm with,}\;\; \tilde{H}_{0} &=& \frac{p_x^2}{2m} + E^{\rm RW}_{+} + E^{\rm RW}_{-}\,\tau_z\,, \non\\
\tilde{H}_{\text{R}} &=& -\frac{\alpha}{\hbar} p_x \tilde{\sigma}_z - \delta_{\rm RW}\,\tilde{\sigma}_y \tau_y\,, \non\\
\tilde{H}_{\text{Z}} &=& B_x \tilde{\sigma}_y + B_y \tilde{\sigma}_z\,. \label{eqn:mcn_nanowire_ham_rw}
\eea

\noin The indices $j$ and $s$ in $\psi_{js}$ label the transverse subband ($j = 1,2$) and spin ($s = \ua,\da$) degrees of freedom, respectively. Here, $E^{\rm RW}_{\pm} = (E^{\rm RW}_1 \pm E^{\rm RW}_2)/2$ denote the average and relative transverse subband energies. The Pauli matrices $\tilde{\sigma}_{x,y,z}$ and $\tau_{x,y,z}$ act in the spin space spanned by ${\chi_{\ua}, \chi_{\da}}$ and in the transverse-mode subspace, respectively. The parameter $\delta_{\rm RW}$ characterizes the coupling–induced hybridization between different transverse subbands with opposite spin orientations and is given by

\beq
\delta_{\rm RW} =\!\!\! \int dy\, \phi^{\rm{RW} \dagger}_{1\ua}(y) \left( -i\alpha \frac{\partial}{\partial y}\sigma_x \right) \phi^{\rm RW}_{2\da}(y) = \frac{8\alpha}{3L_y}\,.
\eeq

\noin When $\delta_{\rm RW}=0$, the transverse subbands are completely decoupled and the nanowire behaves as a set of independent single-channel Rashba wires, corresponding to the decoupled multichannel limit. In momentum space, the multichannel nanowire Hamiltonian in \eqn{eqn:mcn_nanowire_ham_rw} takes the form

\beq
\mathcal{H}^{\rm RW}_{\rm MCN} = \sum_{k}\sum_{\substack{j,j^\prime \\ s,s^\prime}} c_{k,j,s}^\dagger\, \tilde{h}_{j j^\prime; s s^\prime}(k)\, c_{k,j^\prime s^\prime}(k)\,, \non
\eeq

\vspace{-0.5cm}

\bea
\tilde{h} (k) &=& \frac{\hbar^2 k^2}{2m} + E^{\rm RW}_{+} + E^{\rm RW}_{-}\,\tau_z -\alpha k \tilde{\sigma}_z - \delta_{\rm RW}\,\tilde{\sigma}_y \tau_y \non\\ 
&+& B_x \tilde{\sigma}_y + B_y \tilde{\sigma}_z
\eea

\noin A similar analysis of the bulk spectrum can be carried out for the multichannel nanowire with rectangular quantum well confinement, yielding results that are qualitatively similar to those obtained for harmonic confinement.

\begin{figure}[b]
    \centering
    \includegraphics[width=1\linewidth]{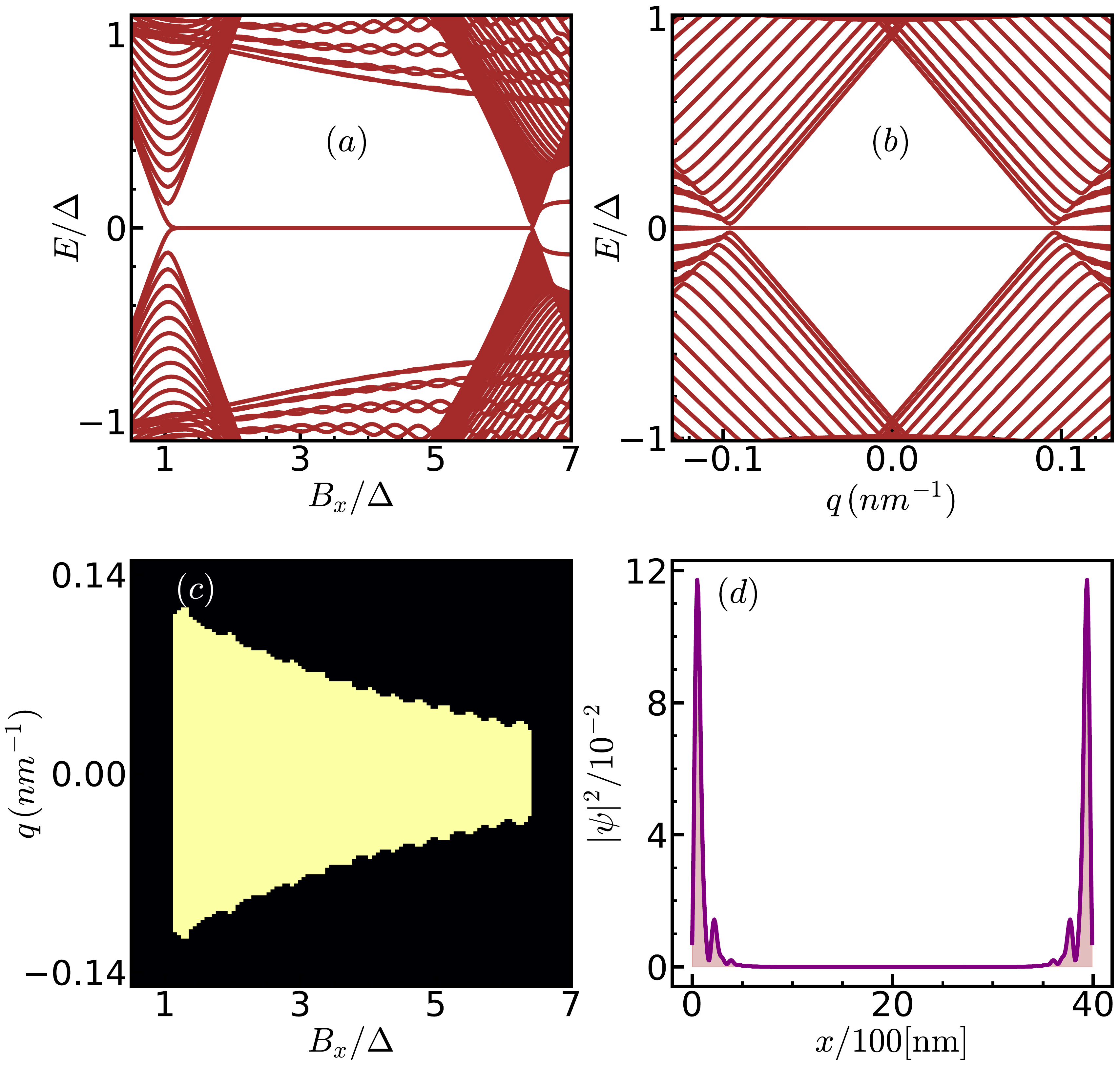}
    \caption{\textbf{Topological signatures of the FF superconducting state for a rectangular well-confined nanowire:} (a) Low-energy spectrum as a function of the longitudinal magnetic field $B_x$, showing zero-energy states over a finite field range for a fixed pairing potential, indicative of Majorana quasiparticles. This regime is broader than that obtained with harmonic confinement. (b) Energy spectrum as a function of finite-momentum pairing $q$ at fixed $B_x / \Delta = 2$. (c) Phase diagram in the $(q, B_x)$ plane, where the yellow region indicates the presence of Majorana zero modes (MZMs) and the black region their absence. (d) Spatial probability density of the Majorana quasiparticle, localized at the edges of the finite lattice. The calculations were performed using $L = 4~\mu\text{m}$, $\Delta = 0.5~\text{meV}$, $\mu = 0.0~\text{meV}$, and $\alpha = 100~\text{meV·nm}$.}
    \label{fig:TSC_rw}
\end{figure}

Following the same self-consistent mean-field protocol employed for the harmonically confined nanowire, we analyze the {FF} superconducting state in the rectangular quantum well geometry. For each value of the Cooper pair momentum $q$, the self-consistency equation \eqn{eqn:self_consistency} is solved to obtain the pairing amplitude $\Delta(q)$. The {FF} ground state is then determined by minimizing the condensation energy with respect to $q$, yielding the optimal Cooper pair momentum $q_0$. As illustrated in \fig{fig:FFLO_state_rw}(a), superconductivity is suppressed beyond critical values of the pairing momentum, occurring either above a positive threshold $q = q_c^+$ or below a negative threshold $q = q_c^-$. Moreover, \fig{fig:FFLO_state_rw}(b) shows that, for a fixed $B_x$, the optimal Cooper pair momentum $q_0$ increases linearly with the transverse Zeeman field $B_y$ in the weak-field regime ($B_y \ll \Delta_0$). Compared to harmonic confinement (\fig{fig:FFLO_state_hc}), the {FF} phase in the rectangular quantum well (\fig{fig:FFLO_state_hc}) exhibits the same qualitative behavior, including well-defined critical momenta $q_c^\pm$ and a linear increase of the optimal momentum $q_0$ with $B_y$ at weak fields. The differences are quantitative: the rectangular confinement slightly reduces the asymmetry, reflecting modified transverse subband spacing and inter-subband coupling.

\subsection{Topologically protected MZMs in the FF superconducting phase} \mylabel{subsec:topology_rectangular_well}

To demonstrate the emergence of topological superconductivity in the multichannel Rashba nanowire with rectangular quantum well confinement, we analyze the quasiparticle spectrum obtained from the lattice Hamiltonian under open boundary conditions, as shown in \fig{fig:TSC_rw}. In this geometry, the lattice Hamiltonian in \eqn{eqn:real_space_ham} is modified by replacing the transverse energy scales $E^{\rm HC}_{\pm}$ with their rectangular-well counterparts $E^{\rm RW}_{\pm}$. Panels \fig{fig:TSC_rw}(a) and \fig{fig:TSC_rw}(b) display the evolution of the low-energy spectrum as functions of the longitudinal Zeeman field $B_x$ and the {FF} momentum $q$. As in the harmonic confinement case, the system undergoes a transition from a trivial to a topological superconducting phase upon increasing $B_x$, signaled by a closing and reopening of the bulk gap and the appearance of zero-energy modes pinned near $E=0$. The current-driven nature of the topological phase persists, with finite-momentum pairing ($q\neq0$) enabling tunability via the applied supercurrent. The corresponding phase diagram shown in \fig{fig:TSC_rw}(c) reveals a significantly enlarged parameter region supporting Majorana zero modes compared to the harmonically confined nanowire, indicating enhanced stability of the topological phase. The real-space probability density in \fig{fig:TSC_rw}(d) confirms the presence of end-localized zero modes, although their spatial localization remains weaker than in the single-channel limit due to inter-subband hybridization.

\begin{figure*}[t]
    \centering
    \includegraphics[width=1\linewidth]{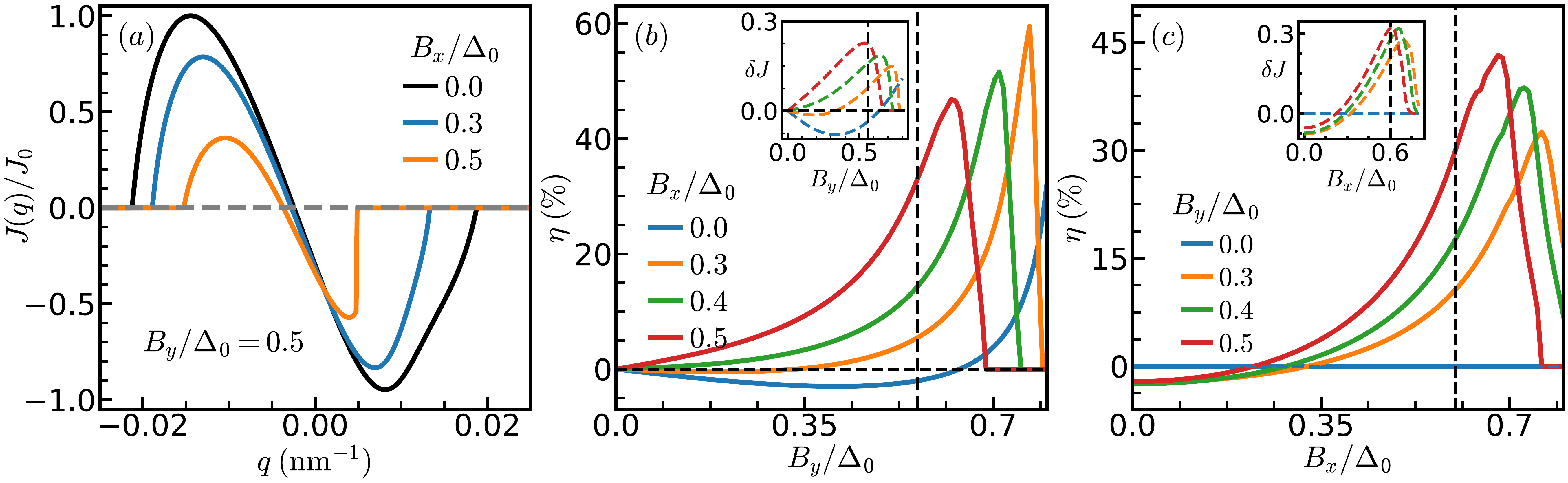}
    \caption{\textbf{Supercurrent density and diode efficiency in a rectangular-well-confined nanowire with interacting channels $(\delta_{\rm RW} \neq 0)$:}  (a) Shows the supercurrent density $J(q)$ as a function of the Cooper-pair momentum $q$ for fixed $B_y / \Delta_0 = 0.5$ and different values of $B_x$. (b,c) Show the corresponding diode efficiency $\eta$ as a function of $B_y / \Delta_0$ for several fixed $B_x$ values (panel b), and as a function of $B_x / \Delta_o$ for several fixed $B_y$ values (panel c).  A finite diode efficiency ${(\eta \neq0)}$ requires both magnetic-field components to be nonzero ($B_y \neq 0$, $B_x \neq 0$).  The model parameters are $(B_z, \alpha, \mu, U, \beta^{-1}) = (0.0, 100~\mathrm{meV\!-\!nm}, 0.0, {15.47}~\mathrm{meV}, 0.1~\mathrm{meV})$.}
    \label{fig:efficiency_interacting_rw}
\end{figure*}

\begin{figure*}[t]
    \centering
    \includegraphics[width=1\linewidth]{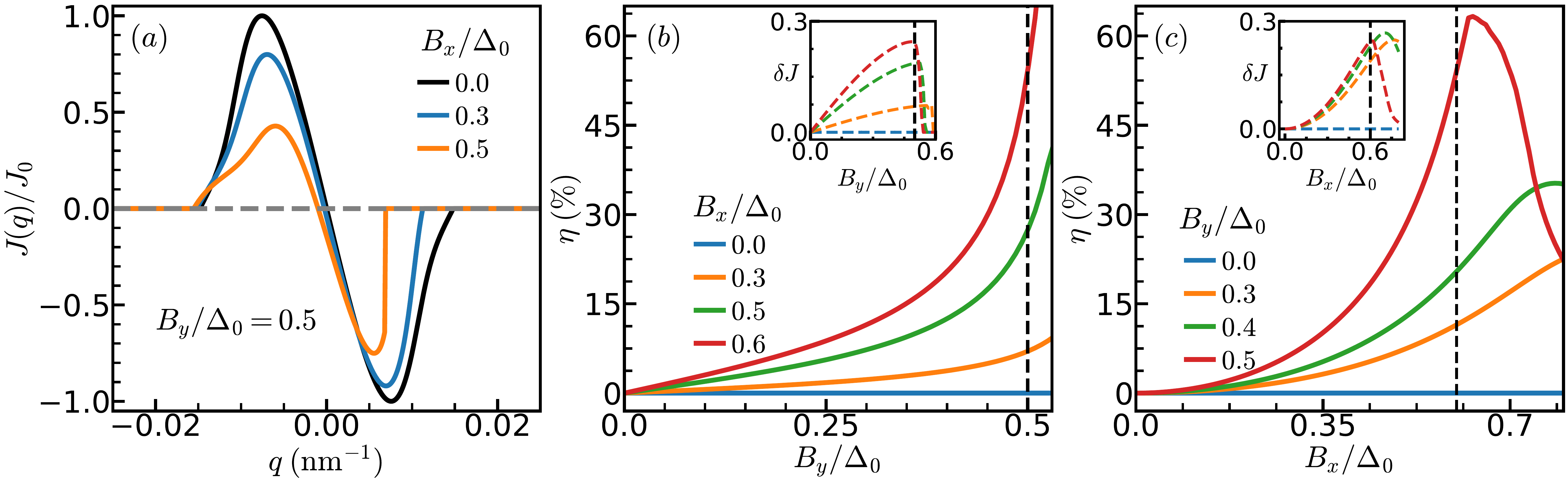}
    \caption{\textbf{Supercurrent density and diode efficiency in a rectangular well-confined nanowire with independent channels $(\delta_{\rm RW} = 0)$:} (a) Supercurrent density $J(q)$ as a function of the Cooper-pair momentum $q$ for fixed $B_y / \Delta_0 = 0.5$ and different values of $B_x$. (b,c) Corresponding diode efficiency $\eta$ as a function of $B_y / \Delta_0$ for several fixed $B_x$ values (panel b), and as a function of $B_x / \Delta_0$ for several fixed $B_y$ values (panel c). A finite diode efficiency is observed even when only a single magnetic field component is present ($B_y \neq 0$). The model parameters are $(B_z, \alpha, \mu, U, \beta^{-1}) = (0.0, 100~\mathrm{meV\!-\!nm}, 0.0, {16}~\mathrm{meV}, 0.1~\mathrm{meV})$.}
    \label{fig:efficiency_independent_rw}
\end{figure*}

\subsection{Nonreciprocal Superconducting Transport} \mylabel{subsec:sde_rectangular_well}

We analyze the superconducting diode effect (SDE) following the same procedure used for evaluating the supercurrent density in Eq. (\ref{SDEcurrent}) and the diode efficiency defined in \eqn{eqn:SDEefficiency} at the same chemical potential ($\mu/\Delta_0 = 0.0$). Both quantities are obtained numerically within a fully self-consistent framework. 

\subsubsection{Case-I: Interacting channels}

The resulting supercurrent density $J(q)$ as a function of the finite Cooper-pair momentum, shown in {Fig.~\ref{fig:efficiency_interacting_rw}(a)}, exhibits a behavior qualitatively similar to that observed in cylindrical nanowires with harmonic confinement, indicating that the fundamental current-carrying mechanism remains preserved in the 2DEG geometry as well. In contrast, the diode efficiency displays a much richer and more non-trivial behavior. As shown in {Fig.~\ref{fig:efficiency_interacting_rw}(b-c)}, the efficiency exhibits sign changes, which reflect a reversal in the preferred direction of supercurrent flow. These sign changes originate from magnetic-field-induced modifications in the electronic band structure, which alter the relative magnitudes of the positive and negative critical currents and thereby strongly influence the degree of supercurrent non-reciprocity. 

Furthermore, it is important to emphasize that a finite diode efficiency persists even when only a single magnetic-field component is present, i.e., for $B_y \neq 0$ and $B_x = 0$. This behavior is consistent with that observed in cylindrical nanowires and highlights the robustness of the SDE in multichannel systems.

\begin{figure}[!ht]
    \centering
    \includegraphics[width=\columnwidth]{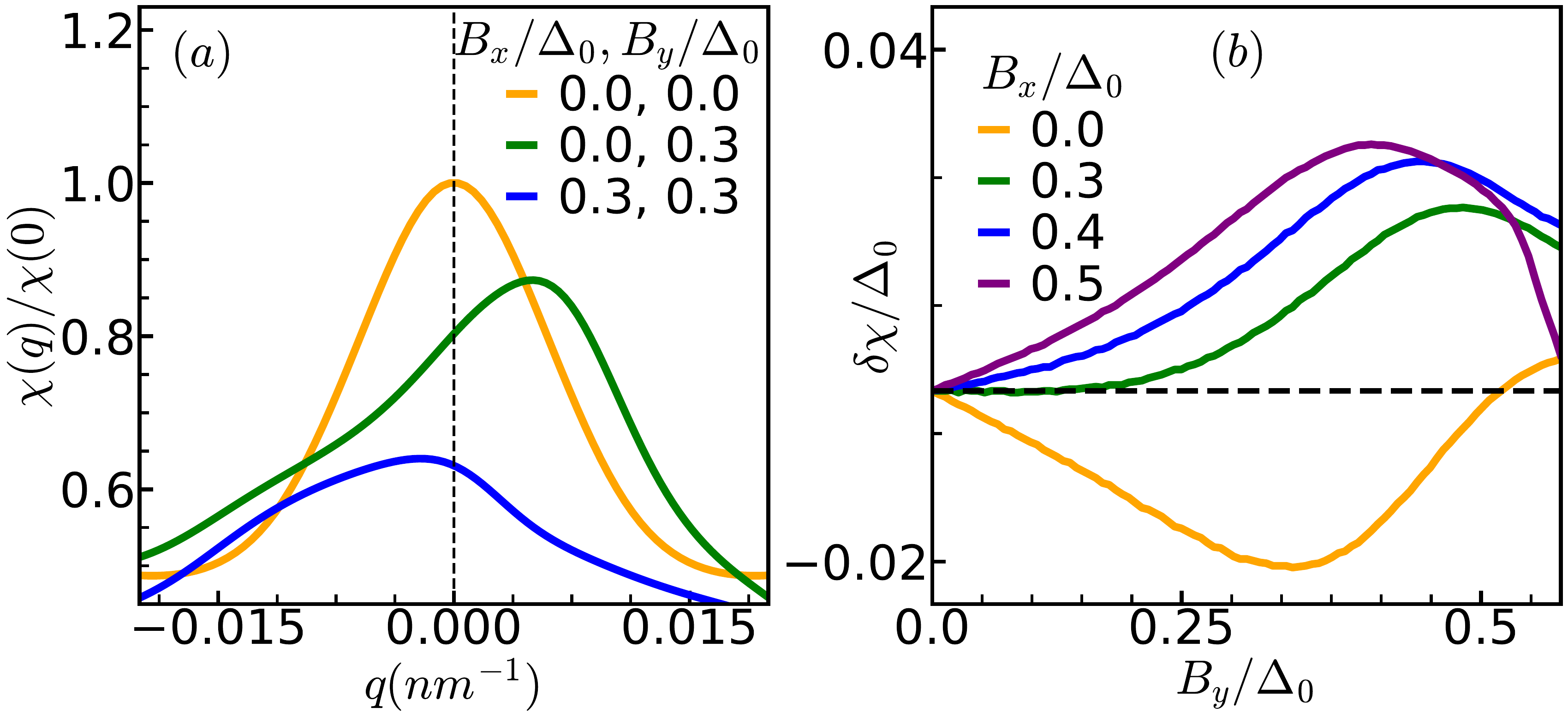}
    \caption{\textbf{Superconducting pairing susceptibility in a rectangular well-confined nanowire:} (a) The superconducting pairing susceptibility is plotted as a function of the Cooper-pair momentum for different values of the transverse magnetic field $B_y$. In contrast to harmonic confinement, the position of the pairing peak shifts with the applied magnetic field, reflecting the magnetic-field-induced modification of the pairing correlations.(b) The difference in pairing susceptibility, evaluated at the Cooper-pair momentum corresponding to the maximum supercurrent, is shown as a function of the transverse magnetic field $B_y$. A clear sign reversal is observed, which is absent in the case of harmonic confinement.}  
    \label{fig:Susceptibility_rw}
\end{figure}

\subsubsection{Case-II: Noninteracting channels}

In contrast, when we consider the case of a \emph{non-interacting channel} in the rectangular quantum well, the overall behavior of the supercurrent density $J(q)$ as a function of the Cooper-pair momentum $q$ remains qualitatively similar to the non-interacting case for cylindrical nanowire. As shown in {Fig.~\ref{fig:efficiency_independent_rw}(a)}, the supercurrent density $J(q)$ is symmetric in $q$ when only a single component of the Zeeman field is present. However, this symmetry is broken once both components of the magnetic field are applied simultaneously, resulting in an asymmetric $J(q)$ profile.

This asymmetry directly gives rise to a finite superconducting diode efficiency $\eta$, as illustrated in {Fig.~\ref{fig:efficiency_independent_rw}(b-c)}. Unlike the interacting case, we observe that the diode efficiency does not exhibit any sign reversal as the magnetic field is varied. Instead, the diode effect remains robust over the entire parameter range considered. Moreover, the magnitude of the diode efficiency is not suppressed in the absence of inter-channel interactions. On the contrary, the effect is slightly enhanced, reaching values as high as $\eta \approx 60\%$. This demonstrates that a sizable superconducting diode effect can be realized even in the non-interacting limit of a rectangular quantum well, highlighting the crucial role played by the combined Zeeman fields and spin--orbit coupling in generating nonreciprocal supercurrent transport.

\subsection{Superconducting Pairing susceptibility in a Rectangular Nanowire} \mylabel{subsec:susceptibility_rectangular_well}

We compute the superconducting pairing susceptibility for the rectangularl well- confined nanowire using the same expression introduced in \eqn{eqn:Susceptibility} for a rectangular nanowire, as shown in \fig{fig:Susceptibility_rw}(a). In the presence of an external magnetic field, the susceptibility develops a pronounced asymmetry, manifested by unequal peak heights at positive and negative pairing momenta, i.e., $\chi(q) \neq \chi(-q)$. This asymmetry signals a directional preference for Cooper pairing and closely mirrors the behavior observed in the harmonically confined nanowire.

In \fig{fig:Susceptibility_rw}(b), we show the difference in the superconducting pairing susceptibility evaluated at the Cooper-pair momenta corresponding to the maximal supercurrents in the forward and reverse directions. As the transverse magnetic field is varied, this difference undergoes a sign change, signaling a reversal in the preferred direction of Cooper pairing over a specific field range. Notably, such behavior is absent in the cylindrically confined nanowire. At larger magnetic fields, the pairing asymmetry is progressively suppressed, revealing a nonmonotonic field dependence that closely tracks the behavior of the superconducting diode efficiency.

\section{Summary and Outlook}  \label{sec:summ}
We have established multichannel Majorana nanowires with distinct transverse confinement geometries as a robust platform for realizing current-driven topological superconductivity and a large SDE. Within a fully self-consistent BdG framework, we analyzed nanowires with harmonic (cylindrical) and rectangular quantum-well confinement and demonstrated that multichannel occupancy generically promotes an asymmetric FF state. This finite-momentum state underlies both the emergence of topological superconductivity with Majorana zero modes over an extended parameter regime and a pronounced nonreciprocal supercurrent response. The externally injected supercurrent directly controls the Cooper-pair momentum, thereby providing a practical tuning knob for the topological transition. For harmonic confinement, we obtain diode efficiencies up to $\sim 60\%$ in the interacting-channel regime and $\sim 55\%$ in the independent-channel limit. Crucially, interchannel coupling qualitatively modifies the symmetry requirements for nonreciprocity, allowing a transverse Zeeman field alone to generate a finite diode response. Rectangular confinement further enhances the effect, yielding efficiencies approaching $\sim 60\%$ in both channel limits and exhibiting a tunable sign reversal of the diode efficiency in the interacting-channel regime. To elucidate the underlying mechanism, we analyzed the superconducting pairing susceptibility $\chi(q)$ and show that a field-induced asymmetry, $\chi(q)\neq\chi(-q)$, stabilizes directional Cooper pairing at intermediate Zeeman fields, thereby providing a microscopic explanation for the nonmonotonic field dependence of the diode response and its suppression at stronger fields. 

A direct comparison of the two confinement geometries reveals both universal and geometry-specific features. In the harmonic case, the diode efficiency reaches approximately $\sim 60\%$ in the interacting-channel regime and about $\sim 55\%$ in the independent-channel limit. The rectangular geometry sustains the topological Fulde–Ferrell phase over a broader parameter range and yields similarly large efficiencies approaching $\sim 60\%$ in both channel limits. Notably, only the rectangular confinement allows a tunable sign reversal of the diode efficiency in the interacting-channel regime, providing a distinct multichannel realization of nonreciprocal superconductivity.

Our results are directly applicable to experimentally accessible spin–orbit–coupled Rashba nanowires, such as zinc-blende InSb and wurtzite InAs, which exhibit strong intrinsic spin–orbit interaction~\cite{Takayanagi1985,liang2012,Mourik2012,das2012,albrecht2016,gul2018}. Nanowires with typical lengths of $\sim100$ nm are well within current fabrication capabilities, and superconductivity can be proximity induced by coupling them to conventional bulk $s$-wave superconductors such as Nb or Al~\cite{Mourik2012,das2012}. In addition, proximity-coupled quasi-one-dimensional strips defined within two-dimensional electron gases offer a highly tunable alternative platform, where the transverse confinement naturally realizes a rectangular quantum well, closely matching the confinement geometry considered here~\cite{annadi2013,Alfonso2023,Settino2020,Udit2019}. An important direction for future work is to examine the robustness of multichannel-induced nonreciprocity against disorder~\cite{Pekerten2017,Pan2024} and interaction effects beyond mean field. Extending these ideas to other spin–orbit–coupled platforms, including nanowires derived from topological insulators, Weyl semimetals, and transition-metal dichalcogenides~\cite{Legg2021ti,Igarashi2017,lim2021nanowire}, as well as nanowire geometries with varying shapes, orientations, or relative misalignment~\cite{Kopasov2021,Christian2018,Kutlin2020}, may further expand the landscape of high-efficiency nonreciprocal superconductivity.

\section{Acknowledgments} 
SKG acknowledges financial support from Anusandhan National Research Foundation (ANRF) erstwhile Science and Engineering Research Board (SERB), Government of India via the Startup Research Grant: SRG/2023/000934 and from IIT Kanpur via the Initiation Grant (IITK/PHY/2022116). The authors utilized the \textit{Andromeda} server at IIT Kanpur for numerical calculations.

\bibliography{refs}

@article{bhowmik_2025,
  title={Optimizing one dimensional superconducting diodes: interplay of Rashba spin-orbit coupling and magnetic fields},
  author={Bhowmik, Sayak and Samanta, Dibyendu and Nandy, Ashis K and Saha, Arijit and Ghosh, Sudeep Kumar},
  journal={Communications Physics},
  volume={8},
  number={1},
  pages={1--8},
  year={2025},
  publisher={Nature Publishing Group},
  url={https://doi.org/10.1038/s42005-025-02044-x}
}

@article{Park_2017,
   title={Andreev spin qubits in multichannel Rashba nanowires},
   volume={96},
   ISSN={2469-9969},
   DOI={10.1103/physrevb.96.125416},
   number={12},
   journal={Physical Review B},
   publisher={American Physical Society (APS)},
   author={Park, Sunghun and Yeyati, A. Levy},
   year={2017},
   month=sep,
   url = {https://doi.org/10.1103/PhysRevB.96.125416}
}

@article{nadeem_2023,
  title={The superconducting diode effect},
  author={Nadeem, Muhammad and Fuhrer, Michael S and Wang, Xiaolin},
  journal={Nature Reviews Physics},
  volume={5},
  number={10},
  pages={558--577},
  year={2023},
  month = sep,
  publisher={Nature Publishing Group UK London},
  url={https://doi.org/10.1038/s42254-023-00632-w}
}

@article{Daido_2022,
  title = {Intrinsic Superconducting Diode Effect},
  author = {Daido, Akito and Ikeda, Yuhei and Yanase, Youichi},
  journal = {Phys. Rev. Lett.},
  volume = {128},
  issue = {3},
  pages = {037001},
  numpages = {6},
  year = {2022},
  month = {Jan},
  publisher = {American Physical Society},
  doi = {10.1103/PhysRevLett.128.037001},
  url = {https://link.aps.org/doi/10.1103/PhysRevLett.128.037001}
}

@article{akito_2022,
  title = {Superconducting diode effect and nonreciprocal transition lines},
  author = {Daido, Akito and Yanase, Youichi},
  journal = {Phys. Rev. B},
  volume = {106},
  issue = {20},
  pages = {205206},
  numpages = {19},
  year = {2022},
  month = {Nov},
  publisher = {American Physical Society},
  doi = {10.1103/PhysRevB.106.205206},
  url = {https://link.aps.org/doi/10.1103/PhysRevB.106.205206}
}

@article{Picoli_2023,
  title = {Superconducting diode effect in quasi-one-dimensional systems},
  author = {de Picoli, Tatiana and Blood, Zane and Lyanda-Geller, Yuli and V\"ayrynen, Jukka I.},
  journal = {Phys. Rev. B},
  volume = {107},
  issue = {22},
  pages = {224518},
  numpages = {6},
  year = {2023},
  month = {Jun},
  publisher = {American Physical Society},
  doi = {10.1103/PhysRevB.107.224518},
  url = {https://link.aps.org/doi/10.1103/PhysRevB.107.224518}
}

@article{Kinnunen_2018,
doi = {10.1088/1361-6633/aaa4ad},
url = {https://dx.doi.org/10.1088/1361-6633/aaa4ad},
year = {2018},
month = {feb},
publisher = {IOP Publishing},
volume = {81},
number = {4},
pages = {046401},
author = {Jami J Kinnunen and Jildou E Baarsma and Jani-Petri Martikainen and Päivi Törmä},
title = {The {Fulde–Ferrell–Larkin–Ovchinnikov} state for ultracold fermions in lattice and harmonic potentials: a review},
journal = {Reports on Progress in Physics},
}

@article{Yuan_2022,
author = {Noah F. Q. Yuan  and Liang Fu },
title = {Supercurrent diode effect and finite-momentum superconductors},
journal = {Proceedings of the National Academy of Sciences},
volume = {119},
number = {15},
pages = {e2119548119},
year = {2022},
doi = {10.1073/pnas.2119548119},
URL = {https://www.pnas.org/doi/abs/10.1073/pnas.2119548119}
}

@article{Legg_2022,
  title = {Superconducting diode effect due to magnetochiral anisotropy in topological insulators and {Rashba} nanowires},
  author = {Legg, Henry F. and Loss, Daniel and Klinovaja, Jelena},
  journal = {Phys. Rev. B},
  volume = {106},
  issue = {10},
  pages = {104501},
  numpages = {8},
  year = {2022},
  month = {Sep},
  publisher = {American Physical Society},
  doi = {10.1103/PhysRevB.106.104501},
  url = {https://link.aps.org/doi/10.1103/PhysRevB.106.104501}
}

@article{Banerjee_2024,
  title = {Enhanced Superconducting Diode Effect due to Coexisting Phases},
  author = {Banerjee, Sayan and Scheurer, Mathias S.},
  journal = {Phys. Rev. Lett.},
  volume = {132},
  issue = {4},
  pages = {046003},
  numpages = {7},
  year = {2024},
  month = {Jan},
  publisher = {American Physical Society},
  doi = {10.1103/PhysRevLett.132.046003},
  url = {https://link.aps.org/doi/10.1103/PhysRevLett.132.046003}
}

@article{ilic_2022,
  title = {Theory of the Supercurrent Diode Effect in Rashba Superconductors with Arbitrary Disorder},
  author = {Ili\ifmmode \acute{c}\else \'{c}\fi{}, S. and Bergeret, F. S.},
  journal = {Phys. Rev. Lett.},
  volume = {128},
  issue = {17},
  pages = {177001},
  numpages = {6},
  year = {2022},
  month = {Apr},
  publisher = {American Physical Society},
  doi = {10.1103/PhysRevLett.128.177001},
  url = {https://link.aps.org/doi/10.1103/PhysRevLett.128.177001}
}

@article{He_2022,
doi = {10.1088/1367-2630/ac6766},
url = {https://dx.doi.org/10.1088/1367-2630/ac6766},
year = {2022},
month = {may},
publisher = {IOP Publishing},
volume = {24},
number = {5},
pages = {053014},
author = {James Jun He and Yukio Tanaka and Naoto Nagaosa},
title = {A phenomenological theory of superconductor diodes},
journal = {New Journal of Physics}
}

@article{Ando2020,
  title={Observation of superconducting diode effect},
  author={Ando, Fuyuki and Miyasaka, Yuta and Li, Tian and Ishizuka, Jun and Arakawa, Tomonori and Shiota, Yoichi and Moriyama, Takahiro and Yanase, Youichi and Ono, Teruo},
  journal={Nature},
  volume={584},
  number={7821},
  pages={373--376},
  year={2020},
  publisher={Nature Publishing Group UK London},
  url = {https://doi.org/10.1038/s41586-020-2590-4}
}

@article{Kopasov2021,
  title = {Geometry controlled superconducting diode and anomalous Josephson effect triggered by the topological phase transition in curved proximitized nanowires},
  author = {Kopasov, A. A. and Kutlin, A. G. and Mel'nikov, A. S.},
  journal = {Phys. Rev. B},
  volume = {103},
  issue = {14},
  pages = {144520},
  numpages = {13},
  year = {2021},
  month = {Apr},
  publisher = {American Physical Society},
  doi = {10.1103/PhysRevB.103.144520},
  url = {https://link.aps.org/doi/10.1103/PhysRevB.103.144520}
}

@article{Kutlin2020,
  title = {Geometry-dependent effects in Majorana nanowires},
  author = {Kutlin, A. G. and Mel'nikov, A. S.},
  journal = {Phys. Rev. B},
  volume = {101},
  issue = {4},
  pages = {045418},
  numpages = {7},
  year = {2020},
  month = {Jan},
  publisher = {American Physical Society},
  doi = {10.1103/PhysRevB.101.045418},
  url = {https://link.aps.org/doi/10.1103/PhysRevB.101.045418}
}

@article{Lutchyn2011,
  title = {Search for Majorana Fermions in Multiband Semiconducting Nanowires},
  author = {Lutchyn, Roman M. and Stanescu, Tudor D. and Das Sarma, S.},
  journal = {Phys. Rev. Lett.},
  volume = {106},
  issue = {12},
  pages = {127001},
  numpages = {4},
  year = {2011},
  month = {Mar},
  publisher = {American Physical Society},
  doi = {10.1103/PhysRevLett.106.127001},
  url = {https://link.aps.org/doi/10.1103/PhysRevLett.106.127001}
}

@article{Liang2012,
  title={Strong tuning of Rashba spin--orbit interaction in single {InAs} nanowires},
  author={Liang, Dong and Gao, Xuan PA},
  journal={Nano Letters},
  volume={12},
  number={6},
  pages={3263--3267},
  year={2012},
  publisher={ACS Publications},
  url = {https://pubs.acs.org/doi/10.1021/nl301325h}
}

@article{Turini2022,
  title={Josephson diode effect in high-mobility InSb nanoflags},
  author={Turini, Bianca and Salimian, Sedighe and Carrega, Matteo and Iorio, Andrea and Strambini, Elia and Giazotto, Francesco and Zannier, Valentina and Sorba, Lucia and Heun, Stefan},
  journal={Nano Letters},
  volume={22},
  number={21},
  pages={8502--8508},
  year={2022},
  publisher={ACS Publications},
  url = {https://doi.org/10.1021/acs.nanolett.2c02899}
}

@article{Takayanagi1985,
  title = {Superconducting Proximity Effect in the Native Inversion Layer on InAs},
  author = {Takayanagi, Hideaki and Kawakami, Tsuyoshi},
  journal = {Phys. Rev. Lett.},
  volume = {54},
  issue = {22},
  pages = {2449--2452},
  numpages = {0},
  year = {1985},
  month = {Jun},
  publisher = {American Physical Society},
  doi = {10.1103/PhysRevLett.54.2449},
  url = {https://link.aps.org/doi/10.1103/PhysRevLett.54.2449}
}

@article{Mourik2012,
author = {V. Mourik  and K. Zuo  and S. M. Frolov  and S. R. Plissard  and E. P. A. M. Bakkers  and L. P. Kouwenhoven },
title = {Signatures of Majorana Fermions in Hybrid Superconductor-Semiconductor Nanowire Devices},
journal = {Science},
volume = {336},
number = {6084},
pages = {1003-1007},
year = {2012},
doi = {10.1126/science.1222360},
URL = {https://www.science.org/doi/abs/10.1126/science.1222360}
}

@article{das2012,
  title={Zero-bias peaks and splitting in an Al--InAs nanowire topological superconductor as a signature of Majorana fermions},
  author={Das, Anindya and Ronen, Yuval and Most, Yonatan and Oreg, Yuval and Heiblum, Moty and Shtrikman, Hadas},
  journal={Nature Physics},
  volume={8},
  number={12},
  pages={887--895},
  year={2012},
  publisher={Nature Publishing Group UK London},
  url={https://doi.org/10.1038/nphys2479}
}

@article{albrecht2016,
  title={Exponential protection of zero modes in Majorana islands},
  author={Albrecht, Sven Marian and Higginbotham, Andrew P and Madsen, Morten and Kuemmeth, Ferdinand and Jespersen, Thomas Sand and Nyg{\aa}rd, Jesper and Krogstrup, Peter and Marcus, CM},
  journal={Nature},
  volume={531},
  number={7593},
  pages={206--209},
  year={2016},
  publisher={Nature Publishing Group UK London},
  url={https://doi.org/10.1038/nature17162}
}

@article{gul2018,
  title={Ballistic Majorana nanowire devices},
  author={G{\"u}l, {\"O}nder and Zhang, Hao and Bommer, Jouri DS and de Moor, Michiel WA and Car, Diana and Plissard, S{\'e}bastien R and Bakkers, Erik PAM and Geresdi, Attila and Watanabe, Kenji and Taniguchi, Takashi and others},
  journal={Nature nanotechnology},
  volume={13},
  number={3},
  pages={192--197},
  year={2018},
  publisher={Nature Publishing Group UK London},
  url={https://doi.org/10.1038/s41565-017-0032-8}
}

@article{Narita_2024,
  author =        {Hideki Narita and Teruo Ono},
  journal =       {JSAP Review},
  number =        {},
  pages =         {240206},
  title =         {Superconducting diode effect in artificial
                   superlattices},
  volume =        {2024},
  year =          {2024},
  doi =           {10.11470/jsaprev.240206},
  url =           {https://doi.org/10.11470/jsaprev.240206},
}

@article{sundaresh_2023,
  author =        {Sundaresh, Ananthesh and V{\"a}yrynen, Jukka I and
                   Lyanda-Geller, Yuli and Rokhinson, Leonid P},
  journal =       {Nature Communications},
  number =        {1},
  pages =         {1628},
  publisher =     {Nature Publishing Group UK London},
  title =         {Diamagnetic mechanism of critical current
                   non-reciprocity in multilayered superconductors},
  volume =        {14},
  year =          {2023},
  url =           {https://doi.org/10.1038/s41467-023-36786-5},
}

@article{Wakatsuki_2017,
  author =        {Ryohei Wakatsuki and Yu Saito and Shintaro Hoshino and
                   Yuki M. Itahashi and Toshiya Ideue and Motohiko Ezawa and
                   Yoshihiro Iwasa and Naoto Nagaosa},
  journal =       {Science Advances},
  number =        {4},
  pages =         {e1602390},
  title =         {Nonreciprocal charge transport in noncentrosymmetric
                   superconductors},
  volume =        {3},
  year =          {2017},
  doi =           {10.1126/sciadv.1602390},
  url =           {https://www.science.org/doi/abs/10.1126/sciadv.1602390},
}

@article{Yuki_2020,
  title={Nonreciprocal transport in gate-induced polar superconductor $\mathrm{SrTiO}_{3}$},
  author={Itahashi, Yuki M and Ideue, Toshiya and Saito, Yu and Shimizu, Sunao and Ouchi, Takumi and Nojima, Tsutomu and Iwasa, Yoshihiro},
  journal={Science advances},
  volume={6},
  number={13},
  pages={eaay9120},
  year={2020},
  publisher={American Association for the Advancement of Science},
  url ={https://www.science.org/doi/abs/10.1126/sciadv.aay9120},
}

@article{Schumann_2020,
  author =        {Schumann, Timo and Galletti, Luca and Jeong, Hanbyeol and
                   Ahadi, Kaveh and Strickland, William M. and
                   Salmani-Rezaie, Salva and Stemmer, Susanne},
  journal =       {Phys. Rev. B},
  month =         {Mar},
  pages =         {100503},
  publisher =     {American Physical Society},
  title =         {Possible signatures of mixed-parity superconductivity
                   in doped polar $\mathrm{SrTiO}_{3}$ films},
  volume =        {101},
  year =          {2020},
  doi =           {10.1103/PhysRevB.101.100503},
  url =           {https://link.aps.org/doi/10.1103/PhysRevB.101.100503},
}

@article{lin_2022,
  author =        {Lin, Jiang-Xiazi and Siriviboon, Phum and
                   Scammell, Harley D and Liu, Song and Rhodes, Daniel and
                   Watanabe, K and Taniguchi, T and Hone, James and
                   Scheurer, Mathias S and Li, JIA},
  journal =       {Nature Physics},
  number =        {10},
  pages =         {1221--1227},
  publisher =     {Nature Publishing Group UK London},
  title =         {Zero-field superconducting diode effect in
                   small-twist-angle trilayer graphene},
  volume =        {18},
  year =          {2022},
  url =           {https://doi.org/10.1038/s41567-022-01700-1},
}

@article{Jaime_2023,
  author =        {Diez-Merida, Jaime and
                   D{\'\i}ez-Carl{\'o}n, Andr{\'e}s and Yang, SY and
                   Xie, Y-M and Gao, X-J and Senior, Jorden and
                   Watanabe, K and Taniguchi, T and Lu, X and
                   Higginbotham, Andrew P and others},
  journal =       {Nature Communications},
  number =        {1},
  pages =         {2396},
  publisher =     {Nature Publishing Group UK London},
  title =         {Symmetry-broken Josephson junctions and
                   superconducting diodes in magic-angle twisted bilayer
                   graphene},
  volume =        {14},
  year =          {2023},
  url =           {https://doi.org/10.1038/s41467-023-38005-7},
}

@article{Chen2025,
  title = {Intrinsic superconducting diode effect and nonreciprocal superconductivity in rhombohedral graphene multilayers},
  author = {Chen, Yinqi and Scheurer, Mathias S. and Schrade, Constantin},
  journal = {Phys. Rev. B},
  volume = {112},
  issue = {6},
  pages = {L060505},
  numpages = {7},
  year = {2025},
  month = {Aug},
  publisher = {American Physical Society},
  doi = {10.1103/zgnk-rw1p},
  url = {https://link.aps.org/doi/10.1103/zgnk-rw1p}
}

@article{bauriedl_2022,
  author =        {Bauriedl, Lorenz and B{\"a}uml, Christian and
                   Fuchs, Lorenz and Baumgartner, Christian and
                   Paulik, Nicolas and Bauer, Jonas M and Lin, Kai-Qiang and
                   Lupton, John M and Taniguchi, Takashi and
                   Watanabe, Kenji and others},
  journal =       {Nature communications},
  number =        {1},
  pages =         {4266},
  publisher =     {Nature Publishing Group UK London},
  title =         {Supercurrent diode effect and magnetochiral
                   anisotropy in few-layer NbSe2},
  volume =        {13},
  year =          {2022},
  url =           {https://doi.org/10.1038/s41467-022-31954-5},
}

@article{Yun_2023,
  author =        {Yun, Jonginn and Son, Suhan and Shin, Jeacheol and
                   Park, Giung and Zhang, Kaixuan and Shin, Young Jae and
                   Park, Je-Geun and Kim, Dohun},
  journal =       {Phys. Rev. Res.},
  month =         {Jun},
  pages =         {L022064},
  publisher =     {American Physical Society},
  title =         {Magnetic proximity-induced superconducting diode
                   effect and infinite magnetoresistance in a van der
                   Waals heterostructure},
  volume =        {5},
  year =          {2023},
  doi =           {10.1103/PhysRevResearch.5.L022064},
  url =           {https://link.aps.org/doi/10.1103/PhysRevResearch.5.L022064},
}

@article{chen2026finite,
  title={Finite-momentum superconductivity from chiral bands in twisted MoTe2},
  author={Chen, Yinqi and Xu, Cheng and Zhang, Yang and Schrade, Constantin},
  journal={Nature Communications},
  year={2026},
  publisher={Nature Publishing Group UK London},
  url={https://doi.org/10.1038/s41467-025-67836-9}
}

@article{Ma2025,
author = {Ma, Jiajun and Zhan, Ruiya and Lin, Xiao},
title = {Superconducting Diode Effects: Mechanisms, Materials and Applications},
journal = {Advanced Physics Research},
volume = {4},
number = {6},
pages = {2400180},
doi = {https://doi.org/10.1002/apxr.202400180},
url = {https://advanced.onlinelibrary.wiley.com/doi/abs/10.1002/apxr.202400180},
year = {2025}
}

@article{shaffer2025,
  title={Theories of superconducting diode effects},
  author={Shaffer, Daniel and Levchenko, Alex},
  journal={arXiv preprint arXiv:2510.25864},
  year={2025},
  url={https://arxiv.org/abs/2510.25864}
}

@article{Zhang_2022,
  title = {General Theory of Josephson Diodes},
  author = {Zhang, Yi and Gu, Yuhao and Li, Pengfei and Hu, Jiangping and Jiang, Kun},
  journal = {Phys. Rev. X},
  volume = {12},
  issue = {4},
  pages = {041013},
  numpages = {11},
  year = {2022},
  month = {Nov},
  publisher = {American Physical Society},
  doi = {10.1103/PhysRevX.12.041013},
  url = {https://link.aps.org/doi/10.1103/PhysRevX.12.041013}
}

@article{Kokkeler_2022,
  title = {Field-free anomalous junction and superconducting diode effect in spin-split superconductor/topological insulator junctions},
  author = {Kokkeler, T. H. and Golubov, A. A. and Bergeret, F. S.},
  journal = {Phys. Rev. B},
  volume = {106},
  issue = {21},
  pages = {214504},
  numpages = {8},
  year = {2022},
  month = {Dec},
  publisher = {American Physical Society},
  doi = {10.1103/PhysRevB.106.214504},
  url = {https://link.aps.org/doi/10.1103/PhysRevB.106.214504}
}

@article{Tanaka_2022,
  title = {Theory of giant diode effect in $d$-wave superconductor junctions on the surface of a topological insulator},
  author = {Tanaka, Yukio and Lu, Bo and Nagaosa, Naoto},
  journal = {Phys. Rev. B},
  volume = {106},
  issue = {21},
  pages = {214524},
  numpages = {13},
  year = {2022},
  month = {Dec},
  publisher = {American Physical Society},
  doi = {10.1103/PhysRevB.106.214524},
  url = {https://link.aps.org/doi/10.1103/PhysRevB.106.214524}
}

@article{Souto_2022,
  title = {Josephson Diode Effect in Supercurrent Interferometers},
  author = {Souto, Rub\'en Seoane and Leijnse, Martin and Schrade, Constantin},
  journal = {Phys. Rev. Lett.},
  volume = {129},
  issue = {26},
  pages = {267702},
  numpages = {6},
  year = {2022},
  month = {Dec},
  publisher = {American Physical Society},
  doi = {10.1103/PhysRevLett.129.267702},
  url = {https://link.aps.org/doi/10.1103/PhysRevLett.129.267702}
}

@article{Cheng_2023,
  title = {Josephson diode based on conventional superconductors and a chiral quantum dot},
  author = {Cheng, Qiang and Sun, Qing-Feng},
  journal = {Phys. Rev. B},
  volume = {107},
  issue = {18},
  pages = {184511},
  numpages = {8},
  year = {2023},
  month = {May},
  publisher = {American Physical Society},
  doi = {10.1103/PhysRevB.107.184511},
  url = {https://link.aps.org/doi/10.1103/PhysRevB.107.184511}
}

@article{Steiner_2023,
  title = {Diode Effects in Current-Biased Josephson Junctions},
  author = {Steiner, Jacob F. and Melischek, Larissa and Trahms, Martina and Franke, Katharina J. and von Oppen, Felix},
  journal = {Phys. Rev. Lett.},
  volume = {130},
  issue = {17},
  pages = {177002},
  numpages = {6},
  year = {2023},
  month = {Apr},
  publisher = {American Physical Society},
  doi = {10.1103/PhysRevLett.130.177002},
  url = {https://link.aps.org/doi/10.1103/PhysRevLett.130.177002}
}

@article{Costa_2023,
  title = {Microscopic study of the Josephson supercurrent diode effect in Josephson junctions based on two-dimensional electron gas},
  author = {Costa, Andreas and Fabian, Jaroslav and Kochan, Denis},
  journal = {Phys. Rev. B},
  volume = {108},
  issue = {5},
  pages = {054522},
  numpages = {13},
  year = {2023},
  month = {Aug},
  publisher = {American Physical Society},
  doi = {10.1103/PhysRevB.108.054522},
  url = {https://link.aps.org/doi/10.1103/PhysRevB.108.054522}
}

@article{Wei_2022,
  title = {Supercurrent rectification effect in graphene-based Josephson junctions},
  author = {Wei, Ya-Jun and Liu, Han-Lin and Wang, J. and Liu, Jun-Feng},
  journal = {Phys. Rev. B},
  volume = {106},
  issue = {16},
  pages = {165419},
  numpages = {7},
  year = {2022},
  month = {Oct},
  publisher = {American Physical Society},
  doi = {10.1103/PhysRevB.106.165419},
  url = {https://link.aps.org/doi/10.1103/PhysRevB.106.165419}
}

@article{Scammell_2022,
doi = {10.1088/2053-1583/ac5b16},
url = {https://dx.doi.org/10.1088/2053-1583/ac5b16},
year = {2022},
month = {mar},
publisher = {IOP Publishing},
volume = {9},
number = {2},
pages = {025027},
author = {Scammell, Harley D and Li, J I A and Scheurer, Mathias S},
title = {Theory of zero-field superconducting diode effect in twisted trilayer graphene},
journal = {2D Materials},
}

@article{Zinkl_2022,
  title = {Symmetry conditions for the superconducting diode effect in chiral superconductors},
  author = {Zinkl, Bastian and Hamamoto, Keita and Sigrist, Manfred},
  journal = {Phys. Rev. Res.},
  volume = {4},
  issue = {3},
  pages = {033167},
  numpages = {12},
  year = {2022},
  month = {Aug},
  publisher = {American Physical Society},
  doi = {10.1103/PhysRevResearch.4.033167},
  url = {https://link.aps.org/doi/10.1103/PhysRevResearch.4.033167}
}

@article{He_2023,
  title={The supercurrent diode effect and nonreciprocal paraconductivity due to the chiral structure of nanotubes},
  author={He, James Jun and Tanaka, Yukio and Nagaosa, Naoto},
  journal={Nature Communications},
  volume={14},
  number={1},
  pages={3330},
  year={2023},
  publisher={Nature Publishing Group UK London},
  url={https://doi.org/10.1038/s41467-023-39083-3}
}

@article{Zhai_2022,
  title = {Prediction of ferroelectric superconductors with reversible superconducting diode effect},
  author = {Zhai, Baoxing and Li, Bohao and Wen, Yao and Wu, Fengcheng and He, Jun},
  journal = {Phys. Rev. B},
  volume = {106},
  issue = {14},
  pages = {L140505},
  numpages = {7},
  year = {2022},
  month = {Oct},
  publisher = {American Physical Society},
  doi = {10.1103/PhysRevB.106.L140505},
  url = {https://link.aps.org/doi/10.1103/PhysRevB.106.L140505}
}

@article{Jiang_2022,
  title = {Field-Free Superconducting Diode in a Magnetically Nanostructured Superconductor},
  author = {Jiang, Ji and Milo\ifmmode \check{s}\else \v{s}\fi{}evi\ifmmode \acute{c}\else \'{c}\fi{}, M.V. and Wang, Yong-Lei and Xiao, Zhi-Li and Peeters, F.M. and Chen, Qing-Hu},
  journal = {Phys. Rev. Appl.},
  volume = {18},
  issue = {3},
  pages = {034064},
  numpages = {9},
  year = {2022},
  month = {Sep},
  publisher = {American Physical Society},
  doi = {10.1103/PhysRevApplied.18.034064},
  url = {https://link.aps.org/doi/10.1103/PhysRevApplied.18.034064}
}

@article{dibyendu_2025,
  title={Field-free Superconducting Diode Effect and Topological Fulde-Ferrell-Larkin-Ovchinnikov Superconductivity in Altermagnetic Shiba Chains},
  author={Samanta, Dibyendu and Ghosh, Sudeep Kumar},
  journal={arXiv:2507.21446},
  year={2025},
  url={https://arxiv.org/abs/2507.21446},
}

@article{bhowmik2025,
  title={Field-free superconducting diode effect in two-dimensional Shiba lattices},
  author={Bhowmik, Sayak and Samanta, Dibyendu and Nandy, Ashis K and Saha, Arijit and Ghosh, Sudeep Kumar},
  journal={arXiv preprint arXiv:2508.10832},
  year={2025},
  url={https://arxiv.org/abs/2508.10832}, 
}

@article{Sayak2025,
  title = {Topological Majorana zero modes and the superconducting diode effect driven by Fulde-Ferrell-Larkin-Ovchinnikov pairing in a helical Shiba chain},
  author = {Bhowmik, Sayak and Saha, Arijit},
  journal = {Phys. Rev. B},
  volume = {111},
  issue = {16},
  pages = {L161402},
  numpages = {6},
  year = {2025},
  month = {Apr},
  publisher = {American Physical Society},
  doi = {10.1103/PhysRevB.111.L161402},
  url = {https://link.aps.org/doi/10.1103/PhysRevB.111.L161402}
}

@article{Amartya2025,
  title={Topological superconductivity and superconducting diode effect mediated via unconventional magnet and Ising spin-orbit coupling},
  author={Pal, Amartya and Mondal, Debashish and Nag, Tanay and Saha, Arijit},
  journal={arXiv preprint arXiv:2512.01266},
  year={2025},
  url={https://arxiv.org/abs/2512.01266}
}

@article{schrade2026,
  title={Altermagnetic superconducting diode effect from non-collinear compensated magnetism in Mn $ \_3 $ Pt},
  author={Schrade, Constantin and Manna, Sujit and Scheurer, Mathias S},
  journal={arXiv preprint arXiv:2601.03348},
  year={2026},
  url={https://arxiv.org/abs/2601.03348}
}

@article{ruthvik2025,
  title={Field-free diode effects in one-dimensional superconductor: a complex interplay between Fulde-Ferrell pairing and altermagnetism},
  author={Ruthvik, SVS and Nag, Tanay},
  journal={arXiv preprint arXiv:2512.01415},
  year={2025},
  url={https://arxiv.org/abs/2512.01415}
}

@article{Fulde1964,
  title = {Superconductivity in a Strong Spin-Exchange Field},
  author = {Fulde, Peter and Ferrell, Richard A.},
  journal = {Phys. Rev.},
  volume = {135},
  issue = {3A},
  pages = {A550--A563},
  numpages = {0},
  year = {1964},
  month = {Aug},
  publisher = {American Physical Society},
  doi = {10.1103/PhysRev.135.A550},
  url = {https://link.aps.org/doi/10.1103/PhysRev.135.A550}
}

@article{ingla_2025,
  title={Efficient superconducting diodes and rectifiers for quantum circuitry},
  author={Ingla-Ayn{\'e}s, Josep and Hou, Yasen and Wang, Sarah and Chu, En-De and Mukhanov, Oleg A and Wei, Peng and Moodera, Jagadeesh S},
  journal={Nature Electronics},
  pages={1--6},
  year={2025},
  publisher={Nature Publishing Group UK London},
  url={https://doi.org/10.1038/s41928-025-01375-5}
}

@article{castellani_2025,
  title={A superconducting full-wave bridge rectifier},
  author={Castellani, Matteo and Medeiros, Owen and Buzzi, Alessandro and Foster, Reed A and Colangelo, Marco and Berggren, Karl K},
  journal={Nature Electronics},
  pages={1--9},
  year={2025},
  publisher={Nature Publishing Group UK London},
  url={https://doi.org/10.1038/s41928-025-01376-4}
}

@article{yuan_2021,
  title={Topological metals and finite-momentum superconductors},
  author={Yuan, Noah FQ and Fu, Liang},
  journal={Proceedings of the National Academy of Sciences},
  volume={118},
  number={3},
  pages={e2019063118},
  year={2021},
  publisher={National Academy of Sciences},
  url = {https://doi.org/10.1073/pnas.2019063118}
}

@article{Sajilesh2025time,
  title={Time-Reversal Symmetry Breaking Superconductivity in HfRhGe: A Noncentrosymmetric Weyl Semimetal},
  author={Sajilesh K, P and Kushwaha, Roshan Kumar and Samanta, Dibyendu and Tula, Tymoteusz and Meena, Pavan Kumar and Srivastava, Shashank and Singh, Deepak and Biswas, Pabitra Kumar and Kanigel, Amit and Hillier, Adrian D and others},
  journal={Advanced Materials},
  volume={37},
  number={7},
  pages={2415721},
  year={2025}
}

@article{Shang2022Weyl,
  title={Spin-triplet superconductivity in {W}eyl nodal-line semimetals},
  author={Shang, Tian and Ghosh, Sudeep K and Smidman, Michael and Gawryluk,
  Dariusz Jakub and Baines, Christopher and Wang, An and Xie, Wu and Chen, Ye and Ajeesh, Mukkattu O and Nicklas, Michael
  and Pomjakushina, Ekaterina and Medarde, Marisa and Shi, Ming and
  Annett, James F. and Yuan, Huiqiu and Quintanilla, Jorge and Shiroka, Toni},
  journal={npj Quantum Mater.},
  volume={7},
  issue={1},
  pages={1--9},
  year={2022},
  publisher={Nature Publishing Group},
  doi = {10.1038/s41535-022-00442-w},
  url={https://doi.org/10.1038/s41535-022-00442-w},
}

@article{Ghosh2020a,
	year = 2020,
	month = {Oct},
	publisher = {{IOP} Publishing},
	volume = {33},
	number = {3},
	pages = {033001},
	author = {S. K. Ghosh and M. Smidman and T. Shang and J. F. Annett and A. D. Hillier and J. Quintanilla and H. Yuan},
	title = {Recent progress on superconductors with time-reversal symmetry breaking},
	journal = {J. Phys. Condens. Matter},
  doi = {10.1088/1361-648X/abaa06},
}

@article{Shang2018,
  title = {Time-Reversal Symmetry Breaking in Re-Based Superconductors},
  author = {Shang, T. and Smidman, M. and Ghosh, S. K. and Baines, C. and Chang, L. J. and Gawryluk, D. J. and Barker, J. A. T. and Singh, R. P. and Paul, D. McK. and Balakrishnan, G. and Pomjakushina, E. and Shi, M. and Medarde, M. and Hillier, A. D. and Yuan, H. Q. and Quintanilla, J. and Mesot, J. and Shiroka, T.},
  journal = {Phys. Rev. Lett.},
  volume = {121},
  issue = {25},
  pages = {257002},
  numpages = {7},
  year = {2018},
  month = {Dec},
  publisher = {American Physical Society},
  doi = {10.1103/PhysRevLett.121.257002},
  url = {https://link.aps.org/doi/10.1103/PhysRevLett.121.257002}
}

@article{Shang2020,
  title = {Time-reversal symmetry breaking in the noncentrosymmetric ${\mathrm{Zr}}_{3}\mathrm{Ir}$ superconductor},
  author = {Shang, T. and Ghosh, S. K. and Zhao, J. Z. and Chang, L.-J. and Baines, C. and Lee, M. K. and Gawryluk, D. J. and Shi, M. and Medarde, M. and Quintanilla, J. and Shiroka, T.},
  journal = {Phys. Rev. B},
  volume = {102},
  issue = {2},
  pages = {020503},
  numpages = {6},
  year = {2020},
  month = {Jul},
  publisher = {American Physical Society},
  doi = {10.1103/PhysRevB.102.020503},
  url = {https://link.aps.org/doi/10.1103/PhysRevB.102.020503}
}

@article{kataria2026,
  title={Observation of Time-Reversal Symmetry Breaking in the Type-I Superconductor YbSb $ \_2$},
  author={Kataria, Anshu and Srivastava, Shashank and Samanta, Dibyendu and Yadav, Pushpendra and Manna, Poulami and Sharma, Suhani and Mishra, Priya and Barker, Joel and Hillier, Adrian D and Agarwal, Amit and others},
  journal={arXiv preprint arXiv:2601.07460},
  year={2026},
  url={https://arxiv.org/abs/2601.07460}
}

@article{Pekerten2017,
  title = {Disorder-induced topological transitions in multichannel Majorana wires},
  author = {Pekerten, B. and Teker, A. and Bozat, \"O. and Wimmer, M. and Adagideli, \ifmmode \dot{I}\else \.{I}\fi{}.},
  journal = {Phys. Rev. B},
  volume = {95},
  issue = {6},
  pages = {064507},
  numpages = {10},
  year = {2017},
  month = {Feb},
  publisher = {American Physical Society},
  doi = {10.1103/PhysRevB.95.064507},
  url = {https://link.aps.org/doi/10.1103/PhysRevB.95.064507}
}

@article{Pan2024,
  title = {Disordered Majorana nanowires: Studying disorder without any disorder},
  author = {Pan, Haining and Das Sarma, Sankar},
  journal = {Phys. Rev. B},
  volume = {110},
  issue = {7},
  pages = {075401},
  numpages = {56},
  year = {2024},
  month = {Aug},
  publisher = {American Physical Society},
  doi = {10.1103/PhysRevB.110.075401},
  url = {https://link.aps.org/doi/10.1103/PhysRevB.110.075401}
}

@article{Legg2021ti,
  title = {Majorana bound states in topological insulators without a vortex},
  author = {Legg, Henry F. and Loss, Daniel and Klinovaja, Jelena},
  journal = {Phys. Rev. B},
  volume = {104},
  issue = {16},
  pages = {165405},
  numpages = {11},
  year = {2021},
  month = {Oct},
  publisher = {American Physical Society},
  doi = {10.1103/PhysRevB.104.165405},
  url = {https://link.aps.org/doi/10.1103/PhysRevB.104.165405}
}

@article{Igarashi2017,
  title = {Magnetotransport in Weyl semimetal nanowires},
  author = {Igarashi, Akira and Koshino, Mikito},
  journal = {Phys. Rev. B},
  volume = {95},
  issue = {19},
  pages = {195306},
  numpages = {8},
  year = {2017},
  month = {May},
  publisher = {American Physical Society},
  doi = {10.1103/PhysRevB.95.195306},
  url = {https://link.aps.org/doi/10.1103/PhysRevB.95.195306}
}

@article{lim2021nanowire,
  title={Nanowire-to-nanoribbon conversion in transition-metal chalcogenides: Implications for one-dimensional electronics and optoelectronics},
  author={Lim, Hong En and Liu, Zheng and Kim, Juan and Pu, Jiang and Shimizu, Hiroshi and Endo, Takahiko and Nakanishi, Yusuke and Takenobu, Taishi and Miyata, Yasumitsu},
  journal={ACS Applied Nano Materials},
  volume={5},
  number={2},
  pages={1775--1782},
  year={2021},
  publisher={ACS Publications},
  url={https://pubs.acs.org/doi/10.1021/acsanm.1c03160}
}

@article{Alfonso2023,
  title = {Hallmarks of orbital-flavored Majorana states in Josephson junctions based on oxide nanochannels},
  author = {Maiellaro, Alfonso and Settino, Jacopo and Guarcello, Claudio and Romeo, Francesco and Citro, Roberta},
  journal = {Phys. Rev. B},
  volume = {107},
  issue = {20},
  pages = {L201405},
  numpages = {6},
  year = {2023},
  month = {May},
  publisher = {American Physical Society},
  doi = {10.1103/PhysRevB.107.L201405},
  url = {https://link.aps.org/doi/10.1103/PhysRevB.107.L201405}
}

@article{Settino2020,
  title = {Spin-orbital polarization of Majorana edge states in oxide nanowires},
  author = {Settino, J. and Forte, F. and Perroni, C. A. and Cataudella, V. and Cuoco, M. and Citro, R.},
  journal = {Phys. Rev. B},
  volume = {102},
  issue = {22},
  pages = {224508},
  numpages = {15},
  year = {2020},
  month = {Dec},
  publisher = {American Physical Society},
  doi = {10.1103/PhysRevB.102.224508},
  url = {https://link.aps.org/doi/10.1103/PhysRevB.102.224508}
}

@article{Udit2019,
  title = {Symmetry and Correlation Effects on Band Structure Explain the Anomalous Transport Properties of (111) ${\mathrm{LaAlO}}_{3}/{\mathrm{SrTiO}}_{3}$},
  author = {Khanna, Udit and Rout, P. K. and Mograbi, Michael and Tuvia, Gal and Leermakers, Inge and Zeitler, Uli and Dagan, Yoram and Goldstein, Moshe},
  journal = {Phys. Rev. Lett.},
  volume = {123},
  issue = {3},
  pages = {036805},
  numpages = {6},
  year = {2019},
  month = {Jul},
  publisher = {American Physical Society},
  doi = {10.1103/PhysRevLett.123.036805},
  url = {https://link.aps.org/doi/10.1103/PhysRevLett.123.036805}
}

@article{annadi2013,
  title={Anisotropic two-dimensional electron gas at the LaAlO3/SrTiO3 (110) interface},
  author={Annadi, A and Zhang, Q and Renshaw Wang, X and Tuzla, Nikolina and Gopinadhan, K and L{\"u}, WM and Roy Barman, A and Liu, ZQ and Srivastava, A and Saha, S and others},
  journal={Nature communications},
  volume={4},
  number={1},
  pages={1838},
  year={2013},
  publisher={Nature Publishing Group UK London},
  url={https://doi.org/10.1038/ncomms2804}
}

@article{Christian2018,
  title = {Geometric Josephson effects in chiral topological nanowires},
  author = {Sp\aa{}nsl\"att, Christian},
  journal = {Phys. Rev. B},
  volume = {98},
  issue = {5},
  pages = {054508},
  numpages = {7},
  year = {2018},
  month = {Aug},
  publisher = {American Physical Society},
  doi = {10.1103/PhysRevB.98.054508},
  url = {https://link.aps.org/doi/10.1103/PhysRevB.98.054508}
}



\end{document}